\documentclass[proof]{WileyASNA-v1}

\usepackage[rightcaption]{sidecap}
\newcommand{\mcitet}[1]{\mbox{\citeauthor{#1}} (\citeyear{#1})}
\newcommand{\mciteauthor}[1]{\mbox{\citeauthor{#1}}}
\newcommand{\mcitealt}[1]{\mbox{\citeauthor{#1}} \citeyear{#1}}
\newcommand{\mcitep}[1]{(\mbox{\citeauthor{#1}} \citeyear{#1})}

\articletype{ORIGINAL ARTICLE}

\received{02 August 2026}
\revised{02 August 2026}
\accepted{13 October 2026}

\begin{document}

\title{The Bochum Survey of the Southern Galactic Disk:\break IV. Multi-Band Photometry}

\author[1]{Julia Blex}

\author[1,2]{Dominik J. Bomans}

\author[1]{Kerstin Weis}

\author[3]{Markus Demleitner}

\author[1]{Moritz Hackstein}

\author[1,2]{\\Anna Franckowiak}

\author[1]{Michael Ramolla}

\author[1]{Christian Westhues}

\author[4]{Christofer Fein}

\author[1,5]{Rolf Chini}

\authormark{BLEX \textsc{et al.}}

\address[1]{\orgdiv{Astronomisches Institut}, \orgname{Ruhr-Universität Bochum}, \orgaddress{\city{Bochum}, \country{Germany}}}

\address[2]{\orgdiv{Ruhr Astroparticle and Plasma Physics (RAPP) Center, \orgaddress{\city{Bochum}, \country{Germany}}}}

\address[3]{\orgdiv{Astronomisches Rechen-Institut}, \orgname{Zentrum für Astronomie der Universität Heidelberg}, \orgaddress{\city{Heidelberg}, \country{Germany}}}

\address[4]{\orgdiv{Clavis Institut für Informationssicherheit}, \orgname{Hochschule Niederrhein}, \orgaddress{\city{Mönchengladbach}, \country{Germany}}}

\address[5]{\orgdiv{Centrum Astronomiczne im. Mikołaja Kopernika}, \orgname{Polskiej Akademii Nauk}, \orgaddress{\city{Warszawa}, \country{Poland}}}

\corres{Julia Blex, Astronomisches Institut, Ruhr-Universität Bochum,\newline Universitätsstraße 150, 44801 Bochum, Germany. \email{julia.blex@rub.de}}

\jnlcitation{\cname{
\author{Blex~J.},
\author{Bomans~D.~J.},
\author{Weis~K.}, 
\author{Demleitner~M.},
\author{Hackstein~M.},
\author{Franckowiak~A.},
\author{Ramolla~M.},
\author{Westhues~C.},
\author{Fein~C.},
\author{Chini~R.}}, (\cyear{2026}),
\ctitle{The Bochum Survey of the Southern Galactic Disk: IV.~Multi-Band Photometry}, \cjournal{Astron. Nachr.}, \cvol{XXXX}.}

\fundingInfo{}

\abstract{The Southern Galactic Disk Survey (GDS) monitored a mosaic of 268~fields along a $6^\circ$-wide stripe in the southern galactic disk from September~2010 to September~2019. These fields were observed between 21~and 407~times simultaneously in $r'$ (\mbox{$8\,\mathrm{mag} \lesssim r' \lesssim 18\,\mathrm{mag}$}) and $i'$ (\mbox{$7\,\mathrm{mag} \lesssim i' \lesssim 17\,\mathrm{mag}$}), supplemented by between 2~and 147~observations in each of the bands $U$, $B$, $V$, and $z'$. Here we present the matched multi-band catalogue, which integrates photometry and light curve data from all bands, containing photometry for almost 21.7~million Gaia~DR3 sources, of which 113\,449 are identified as variables. We give an overview of the survey's variable content with respect to its main characteristics and known variability type classifications. The comparison with the cross-match catalogue of Gaia~DR3 sources with variables from the literature compiled by \mcitet{2023A&A...674A..22G} and the International Variable Star Index (VSX) indicates that approximately $40\,\%$ of the variable sources remain unclassified in terms of variability type. Light curves for all of these sources are contained in the GDS data catalogues, available via the German Astrophysical Virtual Observatory (GAVO), and can contribute to their classification.}


\maketitle

\section{Introduction}
Investigating stellar variability with time-domain observations has advanced stellar astrophysics in many ways (e.g., \mcitealt{2025arXiv251210002H}), such as probing stellar interiors (e.g., \mcitealt{2013ARA&A..51..353C}; \mcitealt{2021RvMP...93a5001A}; \mcitealt{2022ARA&A..60...31K}) and providing insights into the late stages of stellar evolution (e.g., \mcitealt{2026ARA&A..64..103S}). Parameters exclusive to variable stars, including periods and amplitudes, can be employed to derive physical parameters of the stars (e.g., \mcitealt{2007uvs..book.....P}). One example is the period-luminosity relation of Cepheids (\mcitealt{1908AnHar..60...87L}), using which Cepheids can serve as standard candles for estimating distances. Similarly, stellar masses and radii can be determined directly and accurately from eclipsing binary systems (e.g., \mcitealt{2010A&ARv..18...67T}). In turn, distributions of binary properties, such as mass ratios and periods, are valuable for the investigation of star formation (e.g., \mcitealt{2014ApJS..215...15S}), since particular models for the formation of massive stars differ in their assumed properties of the binary and multiple systems (e.g., \mcitealt{2007ARA&A..45..481Z}).

Although stellar variability has been studied by dedicated surveys, among others, the ASAS \mcitep{1997AcA....47..467P}, the \mbox{ASAS-SN} \mcitep{2014ApJ...788...48S}, the OGLE \mcitep{2008AcA....58...69U}, and the ZTF \mcitep{2019PASP..131a8002B}, over the last few decades, there remains a continued need for observation, classification, and characterisation of variable stars in many areas, of which we only mention selected aspects. Regarding the evolution of massive stars, \mcitet{2026arXiv260121530S} points out that current evolutionary models convey an incomplete picture of massive star evolution and that some observables in this context are more complex than anticipated, giving misclassifications of binary interaction products as single stars as an example, and underlines the value of complementary observations, including time-domain photometry for enabling the identification of eclipsing systems. Light curves that extend over several years, as recorded by the Bochum Galactic Disk Survey (GDS), have the potential to contribute to the classification of Luminous Blue Variables (LBVs) of which only a few are known, since to definitely classify LBVs, the observation of at least one SDOR cycle lasting from years to decades (\mcitealt{1997A&AS..124..517V}; \citeyear{1997A&A...318...81V}) or the observation of a giant eruption is necessary \mcitep{2020Galax...8...20W}. Continuous monitoring of variable sources allows to expand the time baseline for the observation of long-term changes like period changes in pulsating variable stars (e.g., \mcitealt{2016JAVSO..44..179N}). Among others, the rate of period change is used as a parameter for studying Cepheid properties (e.g., \mcitealt{2006PASP..118..410T}; \mcitealt{2007PASP..119...82B}; \mcitealt{2012ApJ...760L..18N}; \mcitealt{2022MNRAS.511.2125C}) and the need for observations of nearby Cepheids persists in order to investigate the influence of winds, rotation, and other physical processes (\mcitealt{2016JAVSO..44..179N}).

There have been several catalogues dedicated to compiling variable stars and unifying variability type classifications, particularly the General Catalogue of Variable Stars (GCVS; \mcitealt{2017ARep...61...80S}) and the International Variable Star Index (VSX; \mbox{\mcitealt{2006SASS...25...47W}}). Variability types of nearly 46\,000~variable stars were compiled for Gaia~DR2 by \mcitet{2019A&A...623A.110G}, followed by a cross-match catalogue of Gaia~DR3 with variables from the literature by \mcitet{2023A&A...674A..22G}, which includes 4.9~million variable sources. These catalogues serve as training sets for classification of variable sources (\mcitealt{2023A&A...674A..14R}) and were used to locate the variable stars on the galactic colour-absolute magnitude diagram (e.g., \mcitealt{2019A&A...623A.110G}; \mcitealt{2023A&A...674A..22G}; \mcitealt{2023A&A...674A..14R}), contributing to a global description of stellar variability. $77\,592$ variable stars not listed in either the VSX or the catalogue by \mcitet{2023A&A...674A..22G} could be identified among the stars monitored by the GDS over its entire survey period.

The GDS\footnote{\url{https://galacticdisksurvey.space/}} was conducted from September~2010 to September~2019 with the 15-cm Robotic Bochum Twin Telescope (RoBoTT), formerly known as the VYSOS~6 (V6). A mosaic consisting of 268~fields, each measuring \mbox{$2^{\circ}42\mathrm{'}$}, covering a total area of $1323\,\mathrm{deg}^2$ along a $6^\circ$-wide stripe in the southern galactic disk from RA~$6\mathrm{h}40\mathrm{'}$ to $19\mathrm{h}03\mathrm{'}$ and Dec~$-66^{\circ}03\mathrm{'}$ to $0^{\circ}44\mathrm{'}$ was surveyed. The main objective was to identify variable sources; therefore, the fields were monitored with simultaneous observations in $r'$ and $i'$ (within the magnitude ranges of $8\,\mathrm{mag} \lesssim r' \lesssim 18\,\mathrm{mag}$ and $7\,\mathrm{mag} \lesssim i' \lesssim 17\,\mathrm{mag}$) at a cadence of up to once per day, ending with between 21~and 407~observations per field. Between 2~and 147 additional observations per field in each of the bands $U$, $B$, $V$, and $z'$ were taken, which provide colours of the monitored sources. The observation series consist of 9~dithered images with exposure times of $10\,\mathrm{s}$ in $r'i'$ and of $30\,\mathrm{s}$ in $UBVz'$, respectively. Particular fields were also observed using the narrowband filters $\mathrm{[O}\,\mathrm{III]}$, $\mathrm{H}\alpha$, and $\mathrm{[S}\,\mathrm{II]}$, as well as a narrowband filter centred on $645.0\,\mathrm{nm}$ (NB). Further details on the observation procedure, planned with the Observatory Cerro Armazones (OCA; renamed to Observatory Cerro Murphy, OCM) observational scheduler and database (AutoSched), are given in \mciteauthor{2026AN....34770087B} (\citeyear{2026AN....34770087B}; Paper~III).

The data reduction of the raw images, including bias, dark current and flat-field corrections, the combination of single dithered frames, and the resampling of the images onto a finer grid, was performed with the Bochum OCA VYSOS Interactive Pipeline (BOVIP), as described in \mciteauthor{2012AN....333..706H} (\citeyear{2012AN....333..706H}; Paper~I). The reduced images were further processed by the GDS Analysis and Photometry Pipeline (GDS-APP) to construct light curves and to identify variable sources using variability criteria as defined in \mciteauthor{2015AN....336..590H} (\citeyear{2015AN....336..590H}; Paper~II). By the intermediate state of the survey from 2015, 64\,151~sources (Paper~II) met the variability criteria; the number increased to 113\,449~variable sources in the complete data release (Paper~III).

The complete data release, comprising reduced images, photometry, and light curves, was introduced in Paper~III. Depending on the number of usable observations for a given band-field combination, light curves are available in addition to photometry. Because the observation fields overlap, sources within the overlapping coordinate ranges are measured in up to four fields. Assembling a coherent set of light curves and photometry for each source is non-trivial due to magnitude offsets between fields, image resolution, and varying depth limits across the passbands, particularly in crowded fields. To enhance the usability of the GDS services, we supplement the light curve and photometry catalogues with the multi-band catalogue that compiles photometry for each source across all bands and fields in which it was measured; its construction is described in Section~\ref{Sec:Multi_cat}. We use literature compilations of variability type classifications to assign classifications to the GDS sources in Section~\ref{Sec:source_ident}. Moreover, the image catalogue is complemented by calibration frames, raw image frames, and further reduced images containing standard stars as detailed in Section~\ref{Sec:Img_cat}.

\section{GDS Catalogues}
\subsection{Image Catalogues}\label{Sec:Img_cat}

As described in Papers~I and II, regular observations of Landolt's standard star fields \mcitep{2009AJ....137.4186L} were used for the magnitude calibration. In addition, the RoBoTT observational fields RoBoTT\_0.80px\_0853+0038, RoBoTT\_0.80px\_1242+0030, and RoBoTT\_0.80px\_1539+0040 overlap with Landolt's standard star fields, and therefore the standard stars included in those fields could and did also contribute to the standard star field calibration. These observational fields are included in the RoBoTT image service\footnote{\url{https://dc.g-vo.org/RoBoTT}} available via the Virtual Observatory (VO). For further information on the reduced images, input calibration images, and raw light frames of the RoBoTT images, including the GDS images, refer to the documentation provided with the reduced images.

\subsection{Multi-Band Catalogue}\label{Sec:Multi_cat}
The multi-band catalogue compiles the median magnitudes for each source in all available bands from the second release of the light curve and photometry data\footnote{\url{https://dc.g-vo.org/BGDS2-lc}, \url{https://dc.g-vo.org/BGDS2-med}} (DR2; Paper~III). Besides source coordinates, magnitudes, corresponding fluxes, and Gaia~DR3 identifiers, the catalogue provides variability flags and spectral types. For sources that were measured in $r'$ or $i'$, the coordinates are defined as the median RA and DEC from all $r'$ and $i'$ measurements across all fields, because the master images, from which these coordinates are derived, are usually based on the most images in these bands for a given field. Otherwise, the overall median across all available bands is used. In accordance with the photometry catalogue for individual bands, the magnitudes equal to the median values derived from the light curves of each band. In case of objects that were observed in several fields, the median magnitudes specified are the median values over all fields. Sources are flagged as variable if at least one of their $r'$ or $i'$ light curves is classified as variable in the light curve catalogue; the applied variability criteria are defined in Section~5.1 in Paper~II. The method for determining the spectral types is described in Section~\ref{Sec:SpT}.

The Gaia~DR3 catalogue served as the source catalogue for the multi‑band catalogue; the cross-matching procedure employed is detailed in Section~\ref{Sec:Matching}. Conversely, the match results are specified in the light curve and photometry catalogues, which list individual bands, also using the corresponding Gaia~DR3 identifiers. This enables the identification of the light curves on which the median values in the multi-band catalogue are based. The multi-band catalogue contains nearly 21.7~million sources, comprising the Gaia~DR3 sources that were detected in at least one GDS bands. Photometry in the $UBVr'i'z'$ bands was obtained for approximately $3.4\,\%$, $34.2\,\%$, $67.5\,\%$, $89.7\,\%$, $66.0\,\%$, and $50.3\,\%$ of the sources, respectively. The multi-band catalogue \mcitep{blexetal2025} can be accessed through a browser service\footnote{\url{https://dc.g-vo.org/BGDS2-matched}} and is available via the VO. All access options are listed on the overview page\footnote{\url{http://dc.g-vo.org/browse/bgds/l2}} for the DR2 databases.

\subsubsection{\textbf{Cross-Matching} \label{Sec:Matching}}
The objects from each band-field combination were matched with Gaia~DR3 (\mcitealt{2016A&A...595A...1G}; \citeyear{2023A&A...674A...1G}; \mcitealt{2023A&A...674A..32B}; VizieR table \texttt{I/355/gaiadr3}) symmetrically (i.e., excluding the matching of a source from one catalogue to multiple sources in the other one) within a $3''$ radius. Given Gaia's higher resolution and depth relative to the GDS, magnitude limits are required to avoid mismatches with faint Gaia sources that are not measurable by the GDS. To this end, magnitude limits for the Gaia passbands $G_{\mathrm{BP}}$ and $G_{\mathrm{RP}}$ estimated empirically for each GDS passband were taken into account; these magnitude limits are listed in Table~\ref{tab:mag_limits}.

\begin{table}[t]
\caption{\enspace Magnitude limits for the Gaia passbands depending on the GDS passbands used for the cross-match with Gaia~DR3.}\label{tab:mag_limits}
\centering
\begin{tabular}{lccc}
\toprule
\textbf{Band} & $\lambda_{c}$ \textbf{(nm)} & \textbf{FWHM (nm)} & \textbf{Magnitude limit} \\
\midrule
$U$                        & 365 & 52    & $G_{\mathrm{BP}}~-~U~<~1.5$ \\
$B$                        & 445 & 101   & $G_{\mathrm{BP}}~-~B~<~0.7$ \\
$V$                        & 550 & 82    & $G_{\mathrm{RP}}~-~V~<~0.9$ \\
$r'$                       & 626 & 138   & $G_{\mathrm{RP}}~-~r'~<~0.7$ \\
$i'$                       & 767 & 154   & $G_{\mathrm{RP}}~-~i'~<~0.8$ \\
$z'$                       & 910 & 137   & $G_{\mathrm{RP}}~-~z'~<~2.6$ \\
$\mathrm{[O}\,\mathrm{III]}$ & 500.7 & 6.0 & $G_{\mathrm{BP}}~-~[\mathrm{O}\,\mathrm{III}]~<~1.7$ \\
$\mathrm{NB}$              & 645.0 & 6.0 & $G_{\mathrm{RP}}~-~\mathrm{NB}~<~0.9$ \\
$\mathrm{H}\alpha$         & 656.3 & 6.0 & $G_{\mathrm{RP}}~-~\mathrm{H}\alpha~<~1.0$ \\
$\mathrm{[S}\,\mathrm{II]}$  & 672.1 & 6.0 & $G_{\mathrm{RP}}~-~[\mathrm{S}\,\mathrm{II}]~<~1.2$ \\
\bottomrule
\end{tabular}
\begin{tablenotes}
\item Note: The centre wavelengths $\lambda_c$ and FWHM bandwidths of the RoBoTT passbands used for the GDS are adopted from Tables~1 and~2 in \citet{2016SPIE.9911E..2MR}.
\end{tablenotes}
\end{table}

For nearly 400~variable sources, corresponding detections in at least one band were not assigned a Gaia~DR3 identifier due to the magnitude limits. Frequently, these are highly variable stars with amplitudes on the order of several magnitudes. In these cases, the magnitude and distance limits were unsuitable and were therefore not applied. Ambiguous matches, i.e., cases where multiple Gaia~DR3 sources lie at roughly the same distance from a GDS source, that occasionally (about 1~in~5) occurred were resolved manually through comparison of the Gaia~DR3 coordinates with the GDS images. Fewer than 15\,000~sources across all bands were neither classified as variable nor matched to a Gaia~DR3 source. A manual inspection of the images shows that these are often not point sources, but extended sources such as bright regions in the interstellar medium, and are therefore not included in the multi-band catalogue.

The subset of variable GDS sources was matched to 114\,328~Gaia~DR3 sources. A fraction of the GDS sources listed in the multi-band catalogue lie within an angular distance of only a few arcseconds from one another, for a $3''$ threshold this amounts to roughly $1.5\,\%$ of the sources, although they cannot be resolved in the RoBoTT images. In these cases, it cannot be confirmed which source is variable, or whether variable signals from multiple sources are superimposed. To ensure completeness of the known variability type classifications, all Gaia~DR3 sources are retained in the multi-band catalogue. Conversely, to determine the number of newly identified variables, a lower limit of 113\,449 for the number of variables in the GDS, where multiple sources lying within a radius of $3''$ were counted only once, is used.

\subsubsection{\textbf{Spectral Classification} \label{Sec:SpT}}
The multi-band photometry was used to extract colour information that enables the assignment of spectral types. $UBVr'i'z'$ magnitudes, which are available for all fields, can be considered for the derivation of colours. Although we do not aim to determine effective temperatures or employ a method reliant on preceding classifications of GDS stars, the results of \mcitet{2021NewA...8601568J}, who regress effective temperatures $T_{\mathrm{eff}}$ using random forests, are relevant for inferring trends in spectral types. \mcitet{2021NewA...8601568J} used colours derived from, among others, the NUV, $u'g'r'i'$, and $JHK$ bands, which were compiled from SDSS~DR16 \mcitep{2020ApJS..249....3A}, the GALEX All-Sky Survey \mcitep{2017ApJS..230...24B}, and the AllWISE Data Release \mcitep{2014yCat.2328....0C}, as features. $T_{\mathrm{eff}}$ results from DR6~v2 of LAMOST (\mcitealt{2012RAA....12.1197C}; \mcitealt{2015RAA....15.1095L}) and from the APOGEE Stellar Parameters and Abundances Pipeline (ASPCAP; \mcitealt{2020AJ....160..120J}), applied to APOGEE \mcitep{2017AJ....154...94M} data from SDSS~DR16, served as training and testing samples. $J-H$ and $\mathrm{NUV}-u'$ colours turned out to be most important features of their runs using 8~features for the prediction of $T_{\mathrm{eff}}$. It is anticipated that the most relevant colours for spectral classification vary according to the spectral energy distributions of the stars. The sample used by \mcitet{2021NewA...8601568J} is limited to stars with $T_{\mathrm{eff}}$ between $3000\,\mathrm{K}$ and $8500\,\mathrm{K}$. The near-ultraviolet wavelength range can be assumed to be increasingly relevant for bluer stars, and therefore $U-B$ and $B-V$ colours were used to estimate spectral types of the GDS sources.

Photometrically determined spectral types are listed if $UBV$ magnitudes are available for a source and its colours lie in a range that can be unambiguously assigned to a spectral type. The parametrisation $E_{U - B} / E_{B - V} = 0.72 + 0.05 \, E_{B - V}$ from \mbox{\mcitet{1956ApJ...124..367H}} is employed for the reddening path. Intrinsic colours of stars as a function of the spectral type are taken from Tables~1 and~2 in \mcitet{1970A&A.....4..234F}.

Assuming that the majority of the sample consists of main-sequence stars, luminosity class~V was uniformly assigned to all stars. This hypothesis was examined with the GDS stars' positions on the dust-corrected $r'i'$ Hess diagram. Absolute $i'$ magnitudes are derived using Gaia~DR3 parallaxes (VizieR table \texttt{I/355/gaiadr3}), and $r'$ and $i'$ magnitudes are corrected for dust with extinction curves for the Gaia~DR3 sources determined by \mciteauthor{2025Sci...387.1209Z} (\citeyear{2025Sci...387.1209Z}; VO resource \mcitealt{vo:XP_ap_pars2}), as will be detailed in the next work that includes the Hess diagram of the GDS sources. The stars' positions on the Hess diagram suggest that more than a third of the nearly 12.1~million GDS sources, for which $r'$ and $i'$ photometry, parallaxes, and extinction curves are available, are evolved and that the GDS lacks coverage of the less luminous main-sequence section. Therefore, the percentage of main-sequence stars among the GDS sources should be assumed to be substantially lower than what would be expected from considerations relating to the proportion of the stars' life span spent on the main sequence.

Since no subtype ranges are denoted in \mcitet{1970A&A.....4..234F}, the half distance between two consecutive subtypes in the plane defined by $B-V$ and $U-B$ serves as an approximation of the subtype range. The solutions to the dereddening do not yield unique spectral classifications within the spectral ranges bounded by the reddening paths for types A3 and G0; therefore, these spectral types were omitted. Spectral types were estimated for a total of 306\,989~Gaia~DR3 sources.

\begin{SCfigure*}[1.0][h!]
	\caption{\enspace $UBV$ colour--colour diagram of variable GDS sources. Intrinsic colours of the main-sequence line for O5 to M7 (yellow) are drawn from \mcitet{1970A&A.....4..234F}. Red dashed lines correspond to reddening paths parameterised by \mcitet{1956ApJ...124..367H}; reddening paths at A3 and G0 enclose the colour range that does not provide unique solutions for spectral types. 8514~variable GDS sources for which spectral types were assigned are marked in blue, while sources for which a spectral type assignment was not possible, since they do not fall within the range of the main-sequence line, or the spectral type is ambiguous, are greyed out. \label{fig:UBV}}
	\includegraphics[width=1.414\columnwidth]{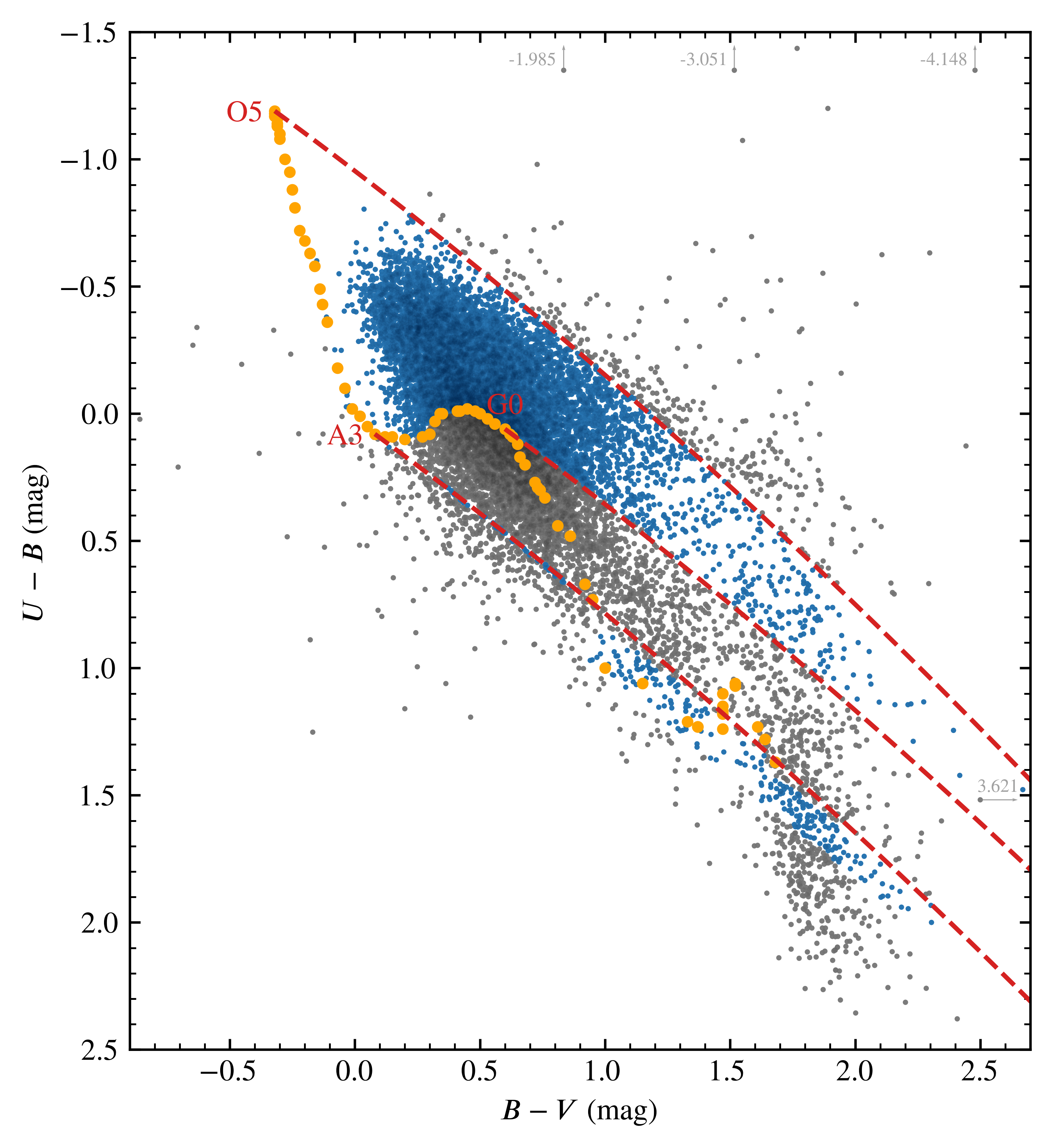}
\end{SCfigure*}

The $UBV$ colour--colour diagram for the subset of variable stars is shown in Figure~\ref{fig:UBV}. The majority of sources lie within the colour range of main-sequence stars. The distribution of spectral types is biased by the lack of $U$ measurements for red sources. A verification of the most distant outliers in the direction of low $U-B$ values using the VSX implies that these sources mainly belong to types of variable stars with high amplitudes, such as Mira variables and semi-regular variables. A time sampling issue is expected to be the cause. Since the magnitudes in $UBV$ rely only on a few observations, the $U$, $B$, and $V$ magnitudes may correspond to different phases, causing large discrepancies for variables with high amplitudes in particular. Nevertheless, this method is limited by the low fraction of sources with photometry in $U$ and is merely sufficient for estimating spectral classes via colour correlations.

\section{Source Identification}\label{Sec:source_ident}
The GDS sources from the multi-band catalogue were identified in the \mbox{SIMBAD} database \mbox{\mcitep{2000A&AS..143....9W}} and in VizieR tables of selected databases, yielding the following percentages of variable GDS sources: $76.1\,\%$ in \mbox{SIMBAD}, $83.2\,\%$ in the VSX (\mcitealt{2006SASS...25...47W}; VizieR table \texttt{B/vsx/vsx}), and $46.0\,\%$ in the compilation of variables from the literature among the Gaia~DR3 sources by \mciteauthor{2023A&A...674A..22G} (\citeyear{2023A&A...674A..22G}, VizieR table \texttt{J/A+A/674/A22/catalog}). $92.0\,\%$ of the objects from the full multi-band catalogue have matches in the 2MASS \mcitep{2006AJ....131.1163S}, $8.8\,\%$ in \mbox{SIMBAD}, and $9.5\,\%$ in the VSX. Given the diversity of source catalogues contributing to the object entries, the cross-matching radius was increased to $8''$ and magnitude limits were not considered for matching against \mbox{SIMBAD} and the VSX. The 2MASS sources were obtained from Gaia~DR3 (VizieR table \texttt{I/355/gaiadr3}), to which the GDS sources have been matched as described in Section~\ref{Sec:Matching}. Gaia~DR3 serves as the source catalogue for both the catalogue by \mcitet{2023A&A...674A..22G} and the multi-band catalogue; therefore, the variable sources from \mcitet{2023A&A...674A..22G} were derived using Gaia~DR3 identifiers. Subsequently, several of these resources will be used to compile stellar properties of the variable stars.

\subsection{Spectral Class}\label{Sec:Ident_SpTs}
Spectral classifications for the multi-band catalogue sources were compiled from \mbox{SIMBAD} and, in a second iteration, spectral classifications from the VSX were incorporated for sources lacking a classification in \mbox{SIMBAD}. Spectral classifications whose subtype was flagged as uncertain were retained, but entries flagged as having an uncertain overall spectral classification were omitted. Binary and multiple star systems comprising components of differing spectral classes were classified according to the spectral class of the primary component; only systems containing WR and OB~stars were assigned to the WR spectral class.

In a third iteration, spectral classes were estimated from Gaia~DR3 $T_{\mathrm{eff}}$ results produced by the Astrophysical parameters inference system (Apsis; \mcitealt{2023A&A...674A..26C}; derived from ARI's Gaia Services\footnote{\url{https://gaia.ari.uni-heidelberg.de}} and VizieR table \texttt{I/355/paramp}). $T_{\mathrm{eff}}$ values for the Gaia~DR3 data were, among others, determined with the General Stellar Parametrizer from Photometry (GSP-Phot) module (\mcitealt{2011MNRAS.411..435B}; \mcitealt{2012MNRAS.426.2463L}; \mcitealt{2023A&A...674A..27A}; Section~11.3.3 of the Gaia~DR3 Documentation\footnote{\url{https://gea.esac.esa.int/archive/documentation/GDR3/Data_analysis/chap_cu8par/sec_cu8par_apsis/ssec_cu8par_apsis_gspphot.html}}) based on BP and RP spectra, $G$ magnitudes, and parallaxes, with the General Stellar Parametrizer from Spectroscopy (GSP-Spec) module (\mcitealt{2023A&A...674A..29R}; Section~11.3.4 of the Gaia~DR3 Documentation\footnote{\url{https://gea.esac.esa.int/archive/documentation/GDR3/Data_analysis/chap_cu8par/sec_cu8par_apsis/ssec_cu8par_apsis_gspspec.html}}) based on Radial Velocity Spectrometer (RVS) spectra, and with the Extended Stellar Parametrizer for Hot Stars (\mbox{ESP-HS}) module (Section~11.3.8 of the Gaia~DR3 Documentation\footnote{\url{https://gea.esac.esa.int/archive/documentation/GDR3/Data_analysis/chap_cu8par/sec_cu8par_apsis/ssec_cu8par_apsis_esphs.html}}) based on BP and RP spectra.

\mcitet{2024ees..conf...59F} compares results for $T_{\mathrm{eff}} > 7500\,\mathrm{K}$ from the GSP-Phot and \mbox{ESP-HS} modules with several reference catalogues, among which the \mbox{LAMOST}~DR5 \mcitep{2019yCat.5164....0L} catalogues and a catalogue of \mbox{O--A}~stars with $T_{\mathrm{eff}} > 7500\,\mathrm{K}$ from \mbox{LAMOST}~DR6 \mcitep{2022A&A...662A..66X} share the largest overlap with the samples of both modules. For A~stars, both modules provide $T_{\mathrm{eff}}$ values with similar bias and spread, while at $T_{\mathrm{eff}} > 10\,000\,\mathrm{K}$, GSP-Phot shows negative biases systematically. \mcitet{2024ees..conf...59F} concludes that $T_{\mathrm{eff}}$ estimates from the GSP-Phot are less reliable for B~stars than those from the \mbox{ESP-HS}. For the O~star range, \mcitet{2024ees..conf...59F} considers galactic and Large Magellanic Cloud stars separately. In the case of galactic stars, which are relevant to the GDS sample, $T_{\mathrm{eff}}$ values from both modules as a function of spectral type from \mbox{SIMBAD} were compared with calibrations between spectral type and $T_{\mathrm{eff}}$ by \mcitet{2010A&A...524A..98W}, showing that the estimates from the \mbox{ESP-HS} lie closer to these calibrations. Therefore, $T_{\mathrm{eff}}$ estimates for the coverage of \mbox{O--A}~stars in the GDS multi-band catalogue were adopted from the \mbox{ESP-HS}. These are supplemented by the GSP-Spec, which mainly provides $T_{\mathrm{eff}}$ results in the \mbox{F--K}~spectral range and is expected to provide more precise values than the GSP-Phot due to the use of RVS~spectra as input data. For the conversion of effective temperatures to spectral types, an updated version\footnote{A Modern Mean Dwarf Stellar Color and Effective Temperature Sequence: \url{http://www.pas.rochester.edu/~emamajek/EEM_dwarf_UBVIJHK_colors_Teff.txt}} of the tables by \mcitet{2013ApJS..208....9P} was used. As for the spectral types derived from GDS $UBV$ photometry in Section~\ref{Sec:SpT}, luminosity class V was assumed for all stars and the half distance between two consecutive spectral classes was set as range for spectral classes.

Because spectral classifications derived from GDS $UBV$ photometry rely solely on photometry rather than spectroscopy, they are expected to be less precise, on average, than those listed in \mbox{SIMBAD} and the VSX, as well as those estimated using Gaia~DR3 effective temperatures. However, since the spectral types from the literature also include photometric classifications from various sources, the precision of those classifications is not necessarily higher in all cases, and spectral classes from GDS $UBV$ photometry benefit from better comparability because they were derived from the same system. The GDS spectral classifications are therefore intended to complement the dataset in cases where no literature spectral classes are available, and are included in all subsequent figures and tables incorporating spectral classes. Spectral types were compiled for $54.9\,\%$ of the variable star candidates in the multi-band catalogue, among them $10\,665$~spectral types from \mbox{SIMBAD}, $24\,925$ from the VSX, 5142 from Gaia~DR3 with $T_{\mathrm{eff}}$ values determined by the \mbox{ESP-HS} module, $19\,907$ from Gaia~DR3 with $T_{\mathrm{eff}}$ values determined by the GSP-Spec module, and 2183 from GDS $UBV$ photometry.

\begin{figure}[t!]
	\centerline{\includegraphics[width=1.0\columnwidth]{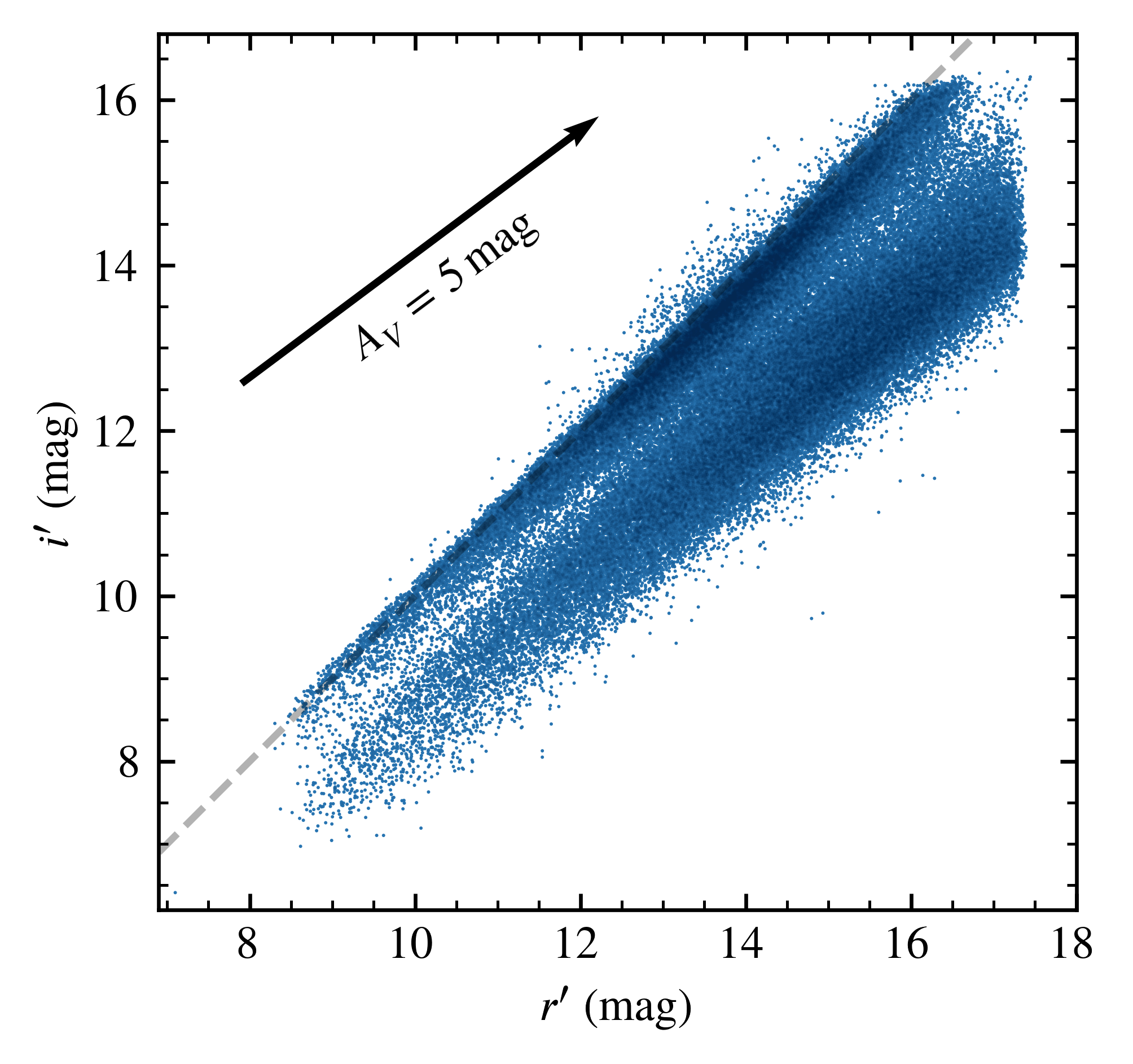}}
	\caption{\enspace $r'$ versus $i'$ of variable sources.\label{fig:GDS_r_i_mag}}
\end{figure}

\begin{figure}[t!]
	\centerline{\includegraphics[width=1.0\columnwidth]{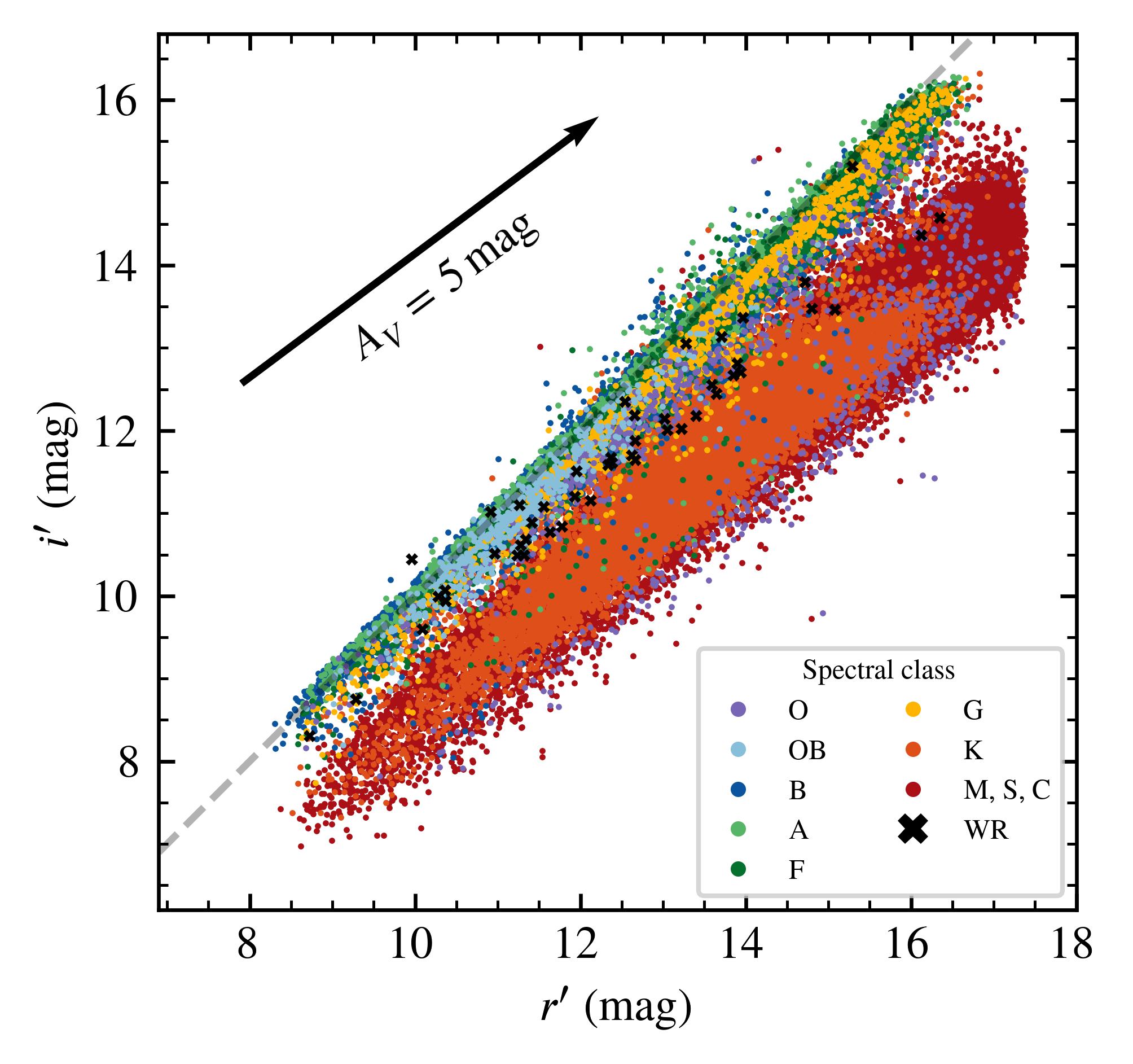}}
	\caption{\enspace $r'$ versus $i'$ of variable stars by spectral class.\label{fig:GDS_r_i_SpT}}
\end{figure}

The relation between the $r'$ and $i'$ magnitudes is shown for all variable GDS sources in Figure~\ref{fig:GDS_r_i_mag}; it is resolved into spectral classes for the subset with available spectral classifications in Figure~\ref{fig:GDS_r_i_SpT}. The parameters for the extinction vectors in Figures~\ref{fig:GDS_r_i_mag} and \ref{fig:GDS_r_i_SpT} were derived from Equations~1--3 in \mcitet{1989ApJ...345..245C}. The relation between $r'$ and $i'$ magnitudes in Figure~\ref{fig:GDS_r_i_mag} shows a bimodality with one subset around the identity line and another consisting of sources brighter in $i'$, nearly parallel to the extinction vector. The relation between the $r'$ and $i'$ magnitudes resolved into spectral classes suggests that the bimodality seen in Figure~\ref{fig:GDS_r_i_mag} originates from the stars' intrinsic colours. Most of the classified \mbox{O--G} stars in the magnitude range covered by the GDS are main-sequence stars, while the majority of the KM stars are evolved, leading to a gap between different luminosity classes.


\subsection{Variability Type}\label{Sec:VarTypes}
We only consider the 114\,328~Gaia~DR3 sources that meet the variability criteria for matching with variable sources from the literature, because their light curves are presumed to offer the most value for further analysis. Since the full magnitude range of the GDS is covered by Gaia~DR3, the cross-match catalogue of Gaia~DR3 sources with variables from the literature by \mcitet{2023A&A...674A..22G} could be used as the primary source for variability type designations for the GDS objects. Only the variability types from the highest-ranked catalogue for each source compiled by \mcitet{2023A&A...674A..22G} were considered. Sources listed as 'OMIT' by \mcitet{2023A&A...674A..22G} are not included in the number of known variables. In addition to variability type classifications from the catalogue by \mcitet{2023A&A...674A..22G}, variability type classifications from the VSX were taken into account for the remaining GDS objects. As with the spectral classifications, variability type classifications marked as uncertain in the VSX were omitted. In total, a variability type could be assigned to approximately $59.7\,\%$ of the variable star candidates from the multi-band catalogue.

In Table~\ref{tab:gds_all_var1}, the variability types are divided into corresponding subtypes; stars exhibiting multiple variability types are listed in each of their main variability groups. For variability types derived from \mcitet{2023A&A...674A..22G}, the classification references sourced from the corresponding VizieR catalogues are listed as footnotes in Table~\ref{tab:gds_all_var1}. Short descriptions of the subtypes are given in Table~\ref{tab:gds_all_var1}; the subtypes are described extensively in Section~4 in \mcitet{2023A&A...674A..22G} and in the variable star type designations in the VSX\footnote{\url{https://vsx.aavso.org/vsx/index.php?view=about.vartypes}}.

In some cases, \mcitet{2023A&A...674A..22G} merged particular subtypes into generic types. Besides additional cPNB[e] and LMXB subtypes, we limit the entries in Table~\ref{tab:gds_all_var1} to those within this classification scheme and incorporate the VSX subtypes into it as follows: UVN (flaring Orion variables) is mapped to FLARES, TTS/ROT (T~Tauri stars showing periodic variability due to spots) to ROT, HADS (high amplitude $\delta$~Scuti variables) and HADS(B) (first/second overtone double-mode $\delta$~Scuti variables) to DSCT, LB (slow irregular variables of late spectral types) and LC (irregular variable supergiants of late spectral types) to L, and IN (Orion variables) and INB (Orion variables of intermediate and late spectral types) to I. Likewise, the convention for assigning subtypes of eclipsing variables to the eclipsing main types EA, EB, EW, and ECL from Table~A.3 in \mcitet{2023A&A...674A..22G} was adopted for additional VSX classifications. Due to the overlap between the definitions of the BE and GCAS ($\gamma$ Cassiopeiae variables) in \mcitet{2023A&A...674A..22G} and the VSX, the BE and GCAS subtypes were merged and are listed under the BE variability type in Table~\ref{tab:gds_all_var1}. \mcitet{2023A&A...674A..22G} and the VSX integrate LBVs into the SDOR variability type; these SDOR classifications were adopted unchanged. For example, $\eta$ Carinae, the name-giving representative \mcitep{1975ApJ...200..426H} of the other type of LBVs besides SDORs, is listed under the SDOR variability type in the VSX.

The division into the extrinsic eclipsing and rotating, as well as the intrinsic pulsating, eruptive, cataclysmic, and X-ray main groups was adopted from the variable star type designations in the VSX. It should be kept in mind that the pulsating and eruptive groups contain many irregular subtypes, whose classifications are expected to be less well-defined than those of periodic variables.

\subsubsection{\textbf{Distribution over Spectral Class}}\label{Sec:VarTypes_SpTs}
\begin{figure}[t]
	\centerline{\includegraphics[width=1.0\columnwidth]{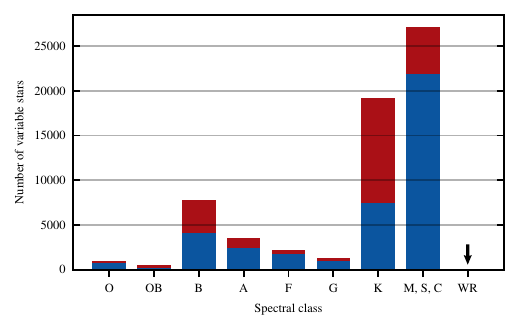}}
	\caption{\enspace Distribution of variable stars with known (blue) and unknown (red) main variability group over spectral class; arrows indicate bars with non-zero counts, i.e., 48~WR~stars.\label{fig:SpT_hist}}
\end{figure}

\begin{figure}[t]
	\centerline{\includegraphics[width=1.0\columnwidth]{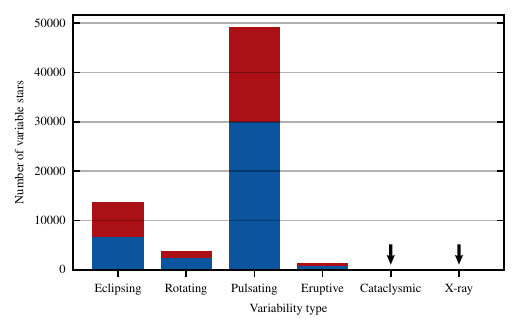}}
	\caption{\enspace Distribution of variable stars with known (blue) and unknown (red) spectral class over main variability group; arrows indicate bars with non-zero counts, i.e., 52~cataclysmic variables and 5~X-ray binaries.\label{fig:var_types_hist}}
\end{figure}

\begin{figure*}[tbh]
\centerline{\includegraphics[width=1.0\textwidth]{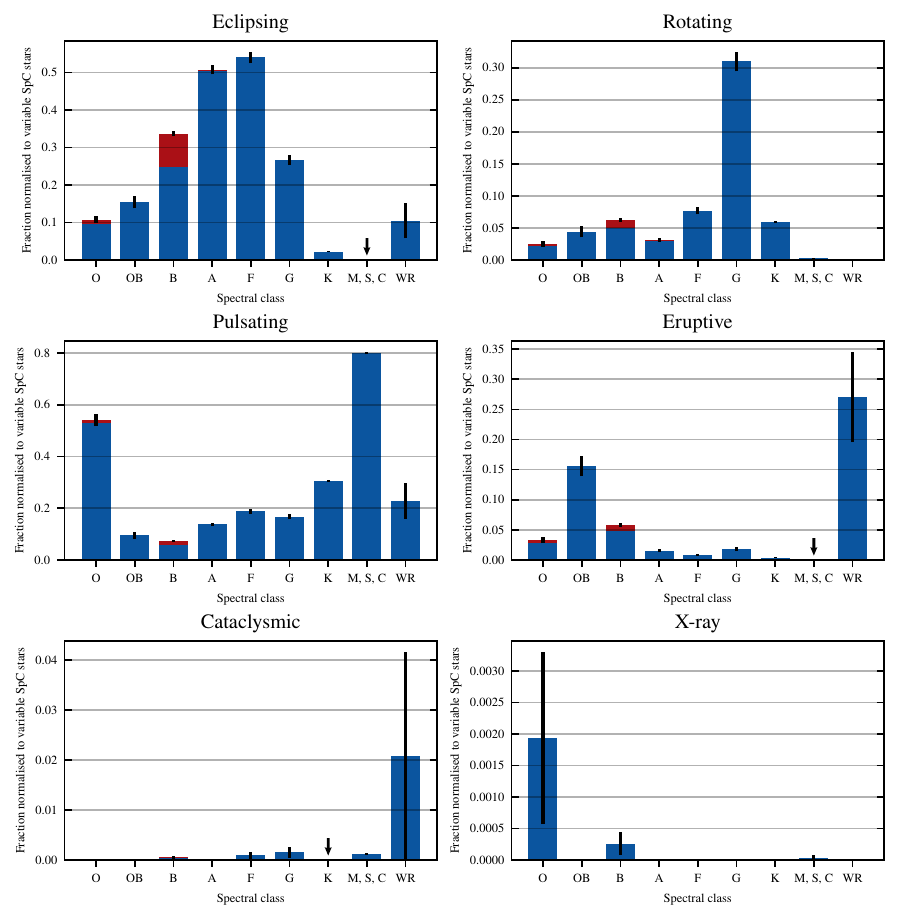}}
\caption{\enspace Fraction of sources in each main variability group, normalised to the variable stars in each spectral class (SpC). Spectral classifications are adopted from \mbox{SIMBAD}, the VSX, and Gaia~DR3 effective temperatures (blue), while additional estimates come from GDS photometry (red). Arrows indicate bars with non-zero counts; error bars are based on the Poisson error. \label{fig:lit_types_hist}}
\end{figure*}

Spectral and variability type classifications for the variable GDS sources are summarised in Table~\ref{tab:gds_types_spt} and Figures~\ref{fig:SpT_hist}--\ref{fig:lit_types_hist}; we examine the spectral distributions of each main variability group shown in Figure~\ref{fig:lit_types_hist} individually.

Since the eclipse probability increases with mass (see Figure~8 in \mcitealt{2005A&A...442.1003S}), the observed decline in eclipse probability from spectral class~F to~O appears implausible. However, because the eclipse probability also depends on inclination, amplitude and orbital period, this distribution cannot be used to determine whether a substantial number of additional eclipsing binaries among the unclassified variables with spectral classes OB might be expected. The GDS provides a homogeneous set of amplitude values, enabling a more comprehensive examination of this trend as detailed in Section~\ref{Sec:EB_Amp}.


Although only convective stars of spectral type F and later exhibit chromospheric activity, a fraction of stars in the \mbox{O--A} spectral range were assigned to the rotating main group. This is due to certain types that involve binarity, which are classified as part of the rotating main group in the VSX. Moreover, LBVs could interfere with the distribution of these spectral classes and Oe and Be stars may exhibit rotation signals through absorption when they possess an asymmetric disk. A manual inspection and period determination of the OB~stars assigned to the rotating main variability group, using the Lafler-Kinman algorithm \mcitep{1965ApJS...11..216L} on ASAS-SN light curves (derived from aperture photometry on the individual images; \mcitealt{2014ApJ...788...48S}; \mcitealt{2017PASP..129j4502K}) and on the GDS light curves, yielded folded light curves whose shapes indicate misclassifications and incorrect periods for some sources. Furthermore, many stars within the ROT subtype compiled by \mcitet{2023A&A...674A..22G} and listed in the VSX, namely the spotted stars among the rotating group according to the VSX definition, were classified based on ASAS-SN light curves (\mcitealt{2019MNRAS.486.1907J}; \citeyear{2019MNRAS.485..961J}; \citeyear{2020MNRAS.491...13J}), using a broader definition of this particular subtype that is not restricted to spotted stars. In fact, rotation can only be reliably verified through long-term spectroscopic observations. Therefore, after further verification of the classifications, a decrease in the number of \mbox{O--A} stars classified as rotating variables might be expected.

The subtypes of pulsating variables occur across different spectral classes (e.g., $\beta$~Cepheids among OB~stars, RR~Lyrae variables among AF~stars, Cepheids among \mbox{F--K}~stars, and Mira variables among M~stars) as largely reflected in the distributions of the particular subtypes.

The distribution of eruptive variables, which shows the highest fractions among massive stars, is notable, because numerous mechanisms causing eruptions, apart from LBVs and other types, occur in convective stars. In particular, the fraction of WR~stars exceeds that found in other spectral classes. Presumably, there are Ofpe/WN9 stars, a spectral type that many LBVs may temporarily exhibit during minimum light \mcitep{1989PASP..101..520B}, among the eruptive variables that have been misclassified as WR~stars.

The class of WR~stars comprises highly heterogeneous subtypes beyond classical WR~stars. Subtypes worth mentioning include WR~stars that are the central stars of planetary nebulae (CSPNe) and very massive WNh stars (for reviews, see \mcitealt{2007ARA&A..45..177C}; \mcitealt{2023A&A...680A..22M}) that are still burning hydrogen (\mcitealt{2008A&A...478..219M}). A cross-match of the variable star candidates from the multi-band catalogue with the catalogue of CSPNe by \mciteauthor{2020A&A...640A..10W} (\citeyear{2020A&A...640A..10W}; VizieR table \texttt{J/A+A/640/A10/catalog}) within an $8''$ radius yields 4~matches, none of which are listed as WR~stars. Although WNh stars can be classical WR~stars in an early evolutionary phase (\mcitealt{2006A&A...457.1015H}; \mcitealt{2014A&A...565A..27H}), they can be used as an indicator for very massive stars among the variable star candidates in the GDS. 7~stars of the WR spectral class in Table~\ref{tab:gds_types_spt} are listed under the WNh sequence, and one further star, also classified as WR variable, is listed under the WN8(h) subtype.

The X-ray variables consist mostly of high-mass X-ray binaries, which are intrinsically hard X-ray emitters and are visible across a large volume of the galactic plane, whereas soft \mbox{X-ray} sources are more prone to absorption by the interstellar medium.

\subsubsection{\textbf{Distribution over Amplitude and Period}}
\begin{figure}[tbh]
	\centerline{\includegraphics[width=1.0\columnwidth]{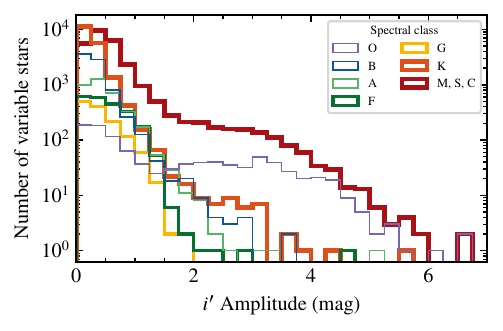}}
	\caption{\enspace $i'$ amplitude distribution of variable stars by spectral class. \label{fig:GDS_i_amp_OBAFGKM}}
\end{figure}

The large number of $r'$ and $i'$ observations, together with strict quality control, which excludes the vast majority of outliers, enables the derivation of reliable amplitude distributions from the corresponding light curves. As demonstrated in Figures~4 and~5 in Paper~II, periods of sources with scatter-corrected amplitudes down to about $0.05\,\mathrm{mag}$ can be determined. Conversely, variations spanning multiple magnitudes can be traced in the light curves (see, e.g., Figure~9 in Paper~III), with the maximum amplitude reaching $10.90\,\mathrm{mag}$. As in Paper~III, the amplitudes were derived from the light curves yielding the highest values for objects observed in several fields. The GDS $i'$ amplitudes were combined with spectral and variability type information in Figures~\ref{fig:GDS_i_amp_OBAFGKM} and~\ref{fig:GDS_i_amp_OBAFGKM_ERP}, and Table~\ref{tab:gds_amp}. The distributions of $r'$ and $i'$ amplitudes do not differ considerably; therefore, the $i'$ amplitudes are shown as an example.

The minimum observable periods in the GDS are limited by the total observation time per observation series. In particular, the $r'$ and $i'$ observation series, consisting of up to 9~images where possible, were observed with an exposure time of $10\,\mathrm{s}$ per frame; however, because of overhead, roughly one minute elapses between individual frame exposures. On the longer end, the observable periods are limited by the time span covered by observations of particular fields, which can exceed 9~years in those monitored over the longest durations.
\mcitet{2023A&A...674A..22G} and the VSX also provide period information for a part of the variable sources, which are compiled in Figure~\ref{fig:GDS_P_OBAFGKM_ERP} and Table~\ref{tab:gds_P}. As with the variability types, periods ranked by \mcitet{2023A&A...674A..22G} were complemented by periods listed in the VSX. Only periods taken from the highest-ranked catalogue in \mcitet{2023A&A...674A..22G} and those from the VSX where no limits or uncertainty flags were specified were included. 

Eclipsing binaries are predominantly B~stars, while pulsations occur more frequently in late-type stars. The amplitude serves as a rough classification criterion; amplitudes higher than $1\,\mathrm{mag}$ rarely originate from rotation, whereas amplitudes exceeding $2.5\,\mathrm{mag}$ are highly likely to be caused by pulsations. The decline in the number of B and A stars with increasing amplitude is plausible, since more mechanisms produce low-amplitude variations. Conversely, the high number of O~stars exhibiting large amplitudes is noteworthy. The period distribution of BA~stars in the eclipsing variability group shown in Figure~\ref{fig:GDS_P_OBAFGKM_ERP} exhibits a clear exponential decline. The periods tend to be longer in pulsating stars than in eclipsing binaries.

The range in periods and amplitudes of the GDS observations in principle allows for the observation of most variability types with few exceptions. These include micro-variability at $\sim\,0.1\,\mathrm{mag}$ of LBVs (\mcitealt{2011NewA...16..421A}) and of evolved massive stars in general (\mcitealt{2019ApJ...878..155D}).

\subsubsection{\textbf{Comparison of Eclipsing Binary Distributions with Simulations}}\label{Sec:EB_Amp}
\mcitet{2005A&A...442.1003S} evolved binary systems to derive eclipse probabilities and distributions over period and eclipse depth. Among others, a time-independent IMF of \mcitet{2001MNRAS.322..231K}, a constant star formation rate, and a duplicity, referring to the percentage of mass fragments from the IMF that are split into doubles, of $80\,\%$ are assumed. \mcitet{2005A&A...442.1003S} confirm that a random pairing model, in which each system component is drawn independently from the IMF, cannot reproduce a sufficient number of binary systems; instead, they use a function with established parameters to describe the distribution of mass ratios between primary and secondary components. A random distribution of inclinations is factored into the eclipse depth $\Delta{m}$.

\begin{figure*}[p]
\centerline{\includegraphics[width=1.00\textwidth]{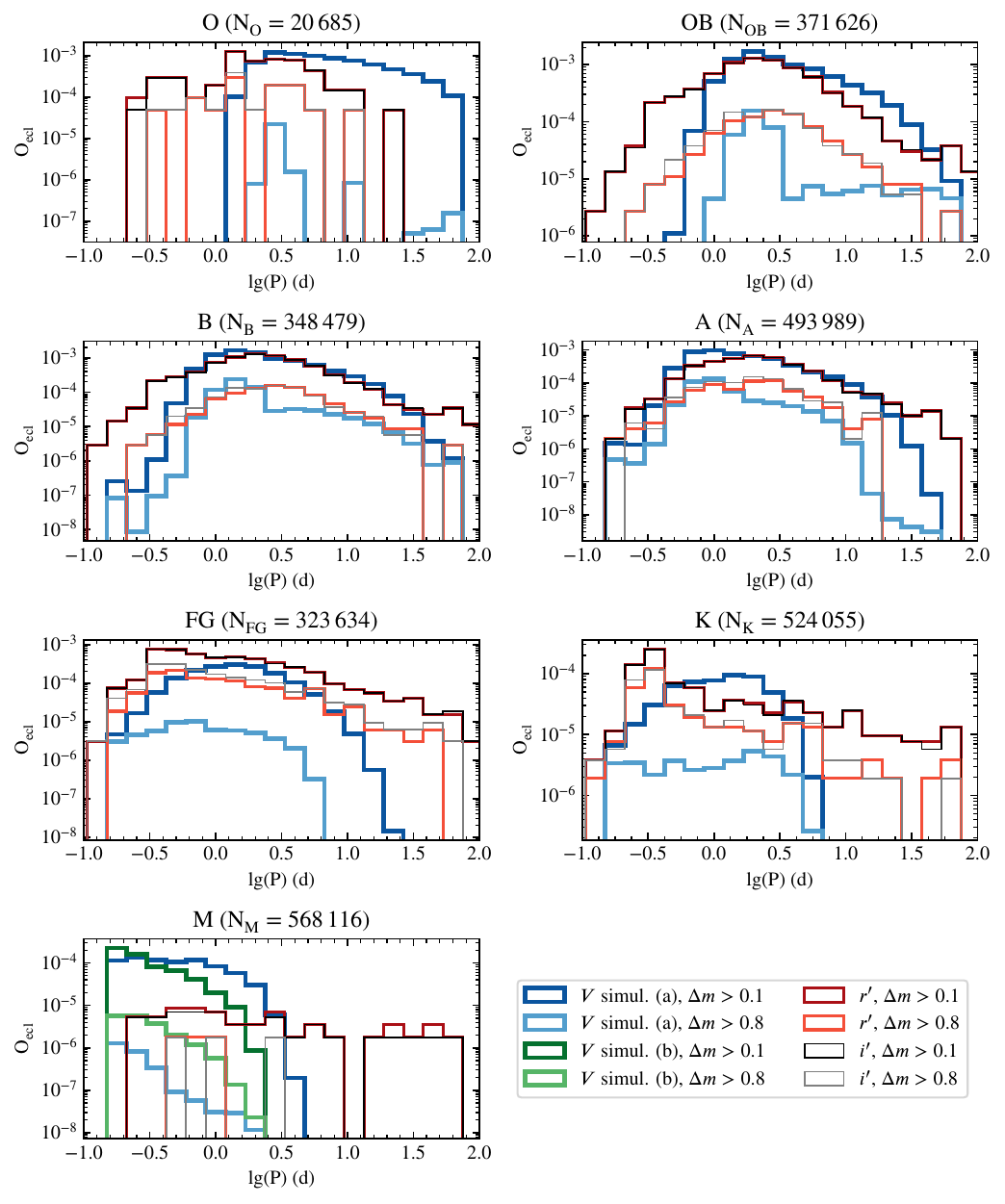}}
\caption{\enspace Distribution of normalised eclipse fractions over period and eclipse depth for main-sequence bins. Simulation-derived eclipse probabilities dependent on $V$ eclipse depths from \mcitet{2005A&A...442.1003S} are compared with observed distributions composed of eclipsing-type classifications and periods from \mcitet{2023A&A...674A..22G} and the VSX, of spectral classifications from \mbox{SIMBAD}, the VSX, Gaia~DR3 effective temperatures, and GDS $UBV$ photometry, and of GDS $r'$ and $i'$ amplitudes. \label{fig:sd_comp}}
\end{figure*}

\mcitet{2005A&A...442.1003S} determine eclipse probabilities for bins of absolute magnitude and colour covering the spectral class range \mbox{O--L} (see Tables~A.1 and~A.2 in that work). In Figure~\ref{fig:sd_comp}, we make use of the compiled periods and $r'$ and $i'$ amplitudes from the eclipsing binary subset for a comparison with these simulations. As in Table~\ref{tab:gds_types_spt}, the spectral classes compiled in Section~\ref{Sec:Ident_SpTs} together with the eclipsing binaries from the catalogue by \mcitet{2023A&A...674A..22G} and the VSX, both of which are expected to be incomplete within the survey's magnitude limit, were used to approximate the number of eclipsing binaries in each spectral class. The normalised eclipsing binary fractions~$\mathrm{O}_{\mathrm{ecl}}$ are derived from the number of variable sources classified as binaries within a particular spectral class that meet the amplitude threshold, divided by the total number of classifications of that spectral class in the multi-band catalogue, as indicated in brackets in the subplots for each spectral class shown in Figure~\ref{fig:sd_comp}.

Eclipses and spectral classifications in the models of \mcitet{2005A&A...442.1003S} differ conceptually from those in the GDS and in the literature compilation; to give a comprehensive account, these discrepancies are specified below.

In the simulations, the probability $P_e(\Delta{m})$ of a primary eclipse deeper than $\Delta{m}$ is counted only if $\Delta{m} > 0.1$ for at least $5\,\%$ of the orbit, to counter the effect of missing observed eclipses as the orbit size increases and eclipses become narrower. In the observed distributions, eclipse depths are replaced by full amplitudes. Eclipse depths from \mcitet{2005A&A...442.1003S} are modelled for relative \mbox{$V$-band} luminosities, whereas amplitudes derived from GDS $V$~observations are unreliable due to the small number of $V$~observations per field, and therefore the $r'$ and $i'$ amplitude distributions were used as earlier.

\mcitet{2005A&A...442.1003S} defined spectral classes by magnitude ranges (see Table~1 in that work), with the “OB” class positioned between the O and B intervals. For the compiled spectral types as listed in Table~\ref{tab:gds_types_spt}, the designation “OB” is exclusively assigned to massive stars whose spectral types could not be specified more precisely, from which separately listed O~and B~stars are distinguished. A spectral classification based on $M_V$ values following the magnitude ranges for the spectral classes defined by \mcitet{2005A&A...442.1003S}, derived from GDS $V$ observations, is not recommended because reddening may affect the results. Hence, the spectral classifications compiled in Section~\ref{Sec:Ident_SpTs} are retained for the distributions shown in Figure~\ref{fig:sd_comp}, except for the “OB” class, which is approximated by the sum of the individual O, B and OB~classifications. Similarly, \mcitet{2005A&A...442.1003S} split the spectral class M into the bins M(a) and M(b), while the compiled spectral classifications in Table~\ref{tab:gds_types_spt} include the entire range of M~subtypes that is covered by the GDS.

To examine the distributions, only the sample with \mbox{${\Delta}m > 0.1$} is considered because of its larger size, which provides more reliable statistics. For the spectral classes \mbox{O--A}, the maximum eclipse probabilities lie at $1\,\mathrm{d} < \mathrm{P} < 10\,\mathrm{d}$, whereas the maximum of low-mass stars lies at $\mathrm{P} < 1\,\mathrm{d}$.

The simulated distributions show fewer short‑period eclipsing binaries (with $\mathrm{P} \lesssim 1\,\mathrm{d}$) for spectral classes \mbox{O--K} than the $r'$ and $i'$ distributions. The discrepancy may be linked to the particular model used for evolving binaries. The simulations employed the rapid binary evolution algorithm by \mcitet{2002MNRAS.329..897H}, which includes magnetic braking, tidal evolution, and Roche‑lobe overflow. Possibly, close binaries evolve more slowly. For example, rapid inward migration of the components (and their eventual merger) can be inhibited by magnetic repulsion, so that they remain for a prolonged time as close binaries with short periods, longer than models predict, which would lead to a higher incidence of such systems.

Likewise, the simulated distributions show fewer long‑period eclipsing binaries (with $\mathrm{P} \gtrsim 10\,\mathrm{d}$) for spectral classes \mbox{A--M} than the $r'$ and $i'$ distributions. A possible explanation is that the simulations work with main-sequence stars only, excluding giants of luminosity classes~\mbox{I--III}, whose radii can be several times larger than those of main-sequence stars. Because the GDS contains giants, it has a higher probability of eclipses even at wider orbits and longer periods. Furthermore, the sample of identified long-period eclipsing binaries is presumably incomplete, as longer periods result in shorter eclipse durations relative to the period, which makes the observation of eclipses less likely. In particular, the deficit of long-period systems among the O~stars compared to the simulated distribution might indicate unknown eclipsing binaries within the O-star population.

\section{Summary and Outlook}
Based on the GDS~DR2 light curves and photometry, we have compiled the multi-band catalogue that contains almost 21.7~million sources, observed in at least one band. The reliable amplitudes measurements have proven useful for statistical studies, such as the investigation of eclipse probabilities.

Among the 114\,328~Gaia~DR3 sources matched to variable GDS sources, the variability types of over 48\,000 \mbox{($\approx\,40\,\%$)} remain unknown, and those with available classifications contain a number of uncertain variability types, which could be worthwhile to examine extensively in the future. Approximately 4200 of the variable stars with pending variability type classifications are presumably OB~stars, including nearly 1100~OB~candidates estimated from GDS $UBV$ photometry. Furthermore, the comparison with normalised eclipse probabilities from \mcitet{2005A&A...442.1003S} might indicate the presence of previously unknown eclipsing binaries among the O~stars. In this regard, the GDS light curves can contribute substantially to the characterisation of unclassified variables and, in particular, OB~stars. Therefore, we hope that the provided light curve and multi-band catalogues will serve as a useful resource for the scientific community.

At present, additional services and resources related to variability are being developed. \mcitet{2023arXiv230403791H} introduced the next generation of the ASAS-SN Sky Patrol service, which expands on the original version that computes individually requested light curves in real time \mcitep{2017PASP..129j4502K}, and provides continuously updated light curves for approximately 111~million targets. Furthermore, \mcitet{2023arXiv230403791H} announced that on this basis, monitoring of specific classes of objects is planned, enabling the examination of extraordinary events in real time. \mcitet{2023A&A...674A..22G} anticipate that the number of variables in the forthcoming Gaia~DR4 will increase by an order of magnitude compared to Gaia~DR3. As mentioned in Paper~III, we are preparing a unified release of the RoBoTT images taken as part of further projects, mostly long-term observations of particular objects, in addition to the GDS. Currently, more than $10\,000$ images are available through the RoBoTT image service. Moreover, future RoBoTT data releases are aimed at implementing enhancements in the data processing pipeline such as adopting current calibration standards (e.g., the standard star databases by \citealt{2026ApJS..282....8H} and \citealt{2026RAA....26e7002L} based on the BEst STars Database by \mcitealt{2026NatSD..13..265X}, released after the suspension of the RoBoTT observations).

We intend to further investigate the content of the GDS in combination with data from the ASAS-SN and Gaia, focusing on eclipsing binaries. Among the high-mass stars, the variable O~stars contained in the GDS have been searched for new eclipsing binaries, yielding 35~new eclipsing systems \mcitep{2019MNRAS.490.5147P}; a double-digit number of stars in the WR spectral sequences and almost 2600~B~stars excluding candidates derived from GDS $UBV$ photometry remain to be investigated. Additionally, more than a thousand of the variable GDS sources are listed as emission-line stars in \mbox{SIMBAD} and might contain further WR star candidates. A provisional revision of the variable light curves revealed numerous previously unknown eclipsing binaries based on the current VSX and VizieR states.

\section*{Acknowledgements}
Standards-compliant publication and browser-based retrieval of the data collections described here is provided by the GAVO data centre, operated by e-inf-astro (BMBF Förderkennzeichen 05A23VH4). D.J.B. and A.F. acknowledge the support from the Deutsche Forschungsgemeinschaft (DFG) via the Collaborative Research Center SFB1491 ‘Cosmic Interacting Matters -- From Source to Signal’. This work was supported by the \mbox{Nordrhein}–Westfälische Akademie der Wissenschaften und der Künste, funded by the Federal State Nordrhein-Westfalen and the Federal Republic of Germany.

This research has made use of the \mbox{SIMBAD} database, operated at CDS, Strasbourg, France.
This research has made use of the VizieR catalogue access tool, CDS, Strasbourg, France (DOI : 10.26093/cds/vizier). The original description of the VizieR service was published in 2000, A{\&}AS 143, 23.
This research made use of the cross-match service provided by CDS, Strasbourg.
This research has made use of the International Variable Star Index (VSX) database, operated at AAVSO, Cambridge, Massachusetts, USA.

This work has made use of data from the European Space Agency (ESA) mission
{\it Gaia} (\url{https://www.cosmos.esa.int/gaia}), processed by the {\it Gaia}
Data Processing and Analysis Consortium (DPAC,
\url{https://www.cosmos.esa.int/web/gaia/dpac/consortium}). Funding for the DPAC
has been provided by national institutions, in particular the institutions
participating in the {\it Gaia} Multilateral Agreement.
This publication makes use of data products from the Two Micron All Sky Survey, which is a joint project of the University of Massachusetts and the Infrared Processing and Analysis Center/California Institute of Technology, funded by the National Aeronautics and Space Administration and the National Science Foundation.

We thank Martin~Haas for providing us with the most recent version and instructions for using the BOVIP. The Tool for OPerations on Catalogues And Tables\footnote{\url{http://www.starlink.ac.uk/topcat/}} (\mbox{TOPCAT}; \mbox{\mcitealt{2005ASPC..347...29T}}) and its command-line counterpart, the Starlink Tables Infrastructure Library Tool Set\footnote{\url{http://www.starlink.ac.uk/stilts/}} (STILTS; \mcitealt{2006ASPC..351..666T}), were used for cross-matching. This work made extensive use of Python together with the packages \texttt{NumPy} \mcitep{harris2020array}, \texttt{pandas} (\mcitealt{reback2020pandas}; \mcitealt{mckinney-proc-scipy-2010}), \texttt{matplotlib} \mcitep{Hunter:2007}, \texttt{seaborn} \mcitep{Waskom2021}, and \texttt{PyEphem}\footnote{\url{https://rhodesmill.org/pyephem/}} \mcitep{2011ascl.soft12014R}. This work made use of Astropy:\footnote{\url{https://www.astropy.org}} a community-developed core Python package and an ecosystem of tools and resources for astronomy (\mcitealt{2013A&A...558A..33A}; \citeyear{2018AJ....156..123A}; \citeyear{2022ApJ...935..167A}).

\subsection*{Conflicts of Interest}
The authors declare no conflicts of interest.

\subsection*{Data Availability Statement}
The GDS data catalogues are publicly available from the GAVO:
\begin{flushleft}
\begin{itemize}
    \item[--] Bochum Galactic Disk Survey images: \\
\url{https://dc.g-vo.org/BGDS}
    \item[--] Robotic Bochum Twin Telescope images: \\
\url{https://doi.org/10.21938/ebLLLoTryChody9S4W1g_g}
    \item[--] Bochum Galactic Disk Survey (BGDS) DR2 light curves: \\
\url{https://doi.org/10.21938/dChYyzuCGA00rsfzfQ8v:Q}
    \item[--] BGDS~DR2 Median Photometry Cone Search: \\
\url{https://doi.org/10.21938/vjJw6sq0gmS5psH9uM5_NA}
    \item[--] BGDS~DR2 Matched Photometry Cone Search: \\
\url{https://doi.org/10.21938/whiHSDPrL:m8cI6AWZ:CFg}
\end{itemize}
\end{flushleft}

\bibliography{GDS_IV}
\appendix
\section{Source Distributions\label{app3}}
\renewcommand{\thetable}{A\arabic{table}}
\onecolumn\begin{sidewaystable}[t]
\caption{\enspace Distribution of variable GDS sources over variability type from the literature.}\label{tab:gds_all_var1}
\centering
\begin{tabular*}{\textheight}{lrl}
\toprule
\textbf{Variability type} & \multicolumn{1}{c}{\textbf{No.}} & \textbf{References} \\
\midrule
\textbf{Eclipsing variability types} & $13754$ & \\
EA ($\beta$ Persei-type (Algol) eclipsing systems) & $5563$ & a (4511), b (449), c (248), d (139), e (101), f (63), g (31), h (13), i (6), \\
& & j (2) \\
EB ($\beta$ Lyrae-type eclipsing systems) & $1939$ & a (1798), c (67), e (55), h (11), d (4), i (3), j (1) \\
EW (W Ursae Majoris-type eclipsing systems) & $2503$ & a (2037), b (131), g (102), d (92), c (64), f (38), e (29), h (7), i (3) \\
ECL (Unspecified eclipsing systems) & $3749$ & d (2651), k (997), e (63), c (31), i (6), l (1) \\
\midrule
\textbf{Rotating variability types} & $3788$ & \\
Chemically peculiar variables & $56$ & \\
\hspace{2mm} -- ACV ($\alpha^2$ Canum Venaticorum variables) & $41$ & m (22), c (15), e (4) \\
\hspace{2mm} -- MCP (Magnetic chemically peculiar variables) & $9$ & n (9) \\
\hspace{2mm} -- CP (Unspecified chemically peculiar variables) & $6$ & h (6) \\
ELL (Ellipsoidal variables) & $61$ & d (25), b (17), c (16), a (3) \\
HB (Heartbeat variables) & $2$ & a (1), d (1) \\
Variables that exhibit chromospheric activity & $3669$ & \\
\hspace{2mm} -- BY (BY Draconis variables) & $58$ & o (38), g (16), c (2), d (2) \\
\hspace{2mm} -- FLARES (Variables that exhibit flares due to magnetic activity) & $3$ & d (3) \\
\hspace{2mm} -- ROT (Unspecified spotted variables) & $2262$ & a (1412), d (589), c (146), q (70), r (29), h (11), s (5) \\
\hspace{2mm} -- BY|ROT (Variables similar to SOLAR\_LIKE including BY and ROT types) & $1$ & p (1) \\
\hspace{2mm} -- RS (RS Canum Venaticorum-type binary systems) & $1341$ & d (1306), g (25), c (7), h (2), t (1) \\
\hspace{2mm} -- UV (UV Ceti variables) & $4$ & d (2), a (1), c (1) \\
\hspace{2mm} -- SOLAR\_LIKE (Unspecified variables that exhibit chromospheric activity) & $1$ & u (1) \\
\midrule
\textbf{Pulsating variability types} & $49169$ & \\
ACYG ($\alpha$ Cygni variables) & $1$ & d (1) \\
BCEP ($\beta$ Cepheids) & $44$ & d (28), c (14), h (1), l (1) \\
\bottomrule
\end{tabular*}
\begin{tablenotes}
\item References: a: \citet{2019MNRAS.485..961J}, b: \citet{2014AcA....64..177S}, c: \citet[version 2019-11-12]{2006SASS...25...47W}, d: \citet[version 2026-06-16]{2006SASS...25...47W}, e: \citet{2002AcA....52..397P}, f: \citet{2008AcA....58..163S}, g: \citet{2020ApJS..249...18C}, h: \citet{2012ApJS..203...32R}, i: Internal\_ECL\_Rybizki, j: \citet{2012A&A...548A..79A}, k: \citet{2018AJ....156..241H}, l: \citet{1997ESASP1200.....E}, m: \citet{2015A&A...581A.138B}, n: \citet{2009A&A...498..961R}, o: Internal\_BY\_Distefano, p: Intenal\_BY\_ROT\_Distefano, q: \citet{2018A&A...618A..30H}, r: \citet{2013A&A...555A..63D}, s: Internal\_ROT\_Distefano, t: \citet{2008MNRAS.389.1722E}, u: \citet{2017ApJ...835...61Z}, v: \citet{2017JKAS...50..131S}, w: \citet{2001A&A...369..178B}, x: \citet{2004AJ....128.2965W}, y: \citet{2019AcA....69..321S}, z: \citet{2017AJ....153..204S}, aa: Internal\_RRL\_Garofalo, ab: \citet{2019A&A...625A..14R}, ac: Internal\_CEP\_Zak, ad: \citet{2019A&A...625A.151B}, ae: \citet{2013A&A...550A.120S}, af: Internal\_DSCT\_DeRidder, ag: \citet{2008A&A...478..659S}, ah: \citet{2020IAUS..345..378V}, ai: \citet{2003A&A...404..301R}, aj: \citet{2019ApJS..240...21A}, ak: \citet{2000A&AS..146..407B}, al: \citet{2019arXiv191205614F}, am: \citet{2019ApJS..244...29M}, an: Internal\_Microlensing\_Kruszy\'nska, ao: \citet{2020ApJS..249...22T}, ap: \citet{2004A&A...424..727P}. The references and labels for internal catalogues from the Gaia Data Processing and Analysis Consortium (DPAC) were drawn from Tab. A.1 in \citet{2023A&A...674A..22G}. VizieR catalogues corresponding to the references can be found in the VizieR table J/A+A/674/A22/notes. The number of sources retrieved from each reference is given in brackets. 
\end{tablenotes}
\end{sidewaystable}
\clearpage
\begin{sidewaystable}[t]
\addtocounter{table}{-1}\caption{\enspace Continued.}\label{tab:gds_all_var2}
\centering
\begin{tabular*}{\textheight}{lrl}
\toprule
\textbf{Variability type} & \multicolumn{1}{c}{\textbf{No.}} & \textbf{References} \\
\midrule
Cepheids & $859$ & \\
\hspace{2mm} -- ACEP (Anomalous Cepheids) & $2$ & b (1), d (1) \\
\hspace{2mm} -- BLHER (BL Herculis variables) & $37$ & a (19), b (15), ab (3) \\
\hspace{2mm} -- CW (W Virginis variables) & $81$ & b (54), a (14), ab (8), d (3), c (2) \\
\hspace{2mm} -- DCEP ($\delta$ Cepheids) & $612$ & b (334), ab (154), e (43), a (36), c (19), q (15), d (9), g (1), l (1) \\
\hspace{2mm} -- RV (RV Tauri variables) & $86$ & b (74), a (5), h (3), c (2), d (1), ab (1) \\
\hspace{2mm} -- T2CEP (Unspecified Type II Cepheids)  & $35$ & q (33), g (1), ad (1) \\
\hspace{2mm} -- CEP (Unspecified Cepheids) & $6$ & c (3), ac (2), j (1) \\
DSCT ($\delta$ Scuti variables) & $238$ & a (190), c (13), d (10), e (10), g (6), h (4), f (3), l (1), ae (1) \\
DSCT|SXPHE ($\delta$ Scuti and SX Phoenicis variables) & $5$ & q (4), af (1) \\
GDOR ($\gamma$ Doradus variables) & $5$ & ae (4), d (1) \\
Long-period and semi-regular variables & $36448$ & \\
\hspace{2mm} -- LSP (Long-secondary-period variables) & $71$ & h (71) \\
\hspace{2mm} -- M (Mira variables) & $5881$ & d (4525), f (832), g (241), c (138), e (93), k (24), x (20), a (6), \\
& & j (1), l (1) \\
\hspace{2mm} -- M|SR (Long-period variables including M and SR types) & $10560$ & d (5677), q (4883) \\
\hspace{2mm} -- SARG (Small amplitude red giants) & $199$ & h (199) \\
\hspace{2mm} -- OSARG (Small amplitude red giants from OGLE) & $491$ & f (491) \\
\hspace{2mm} -- SR (Semi-regular variables) & $8252$ & d (3549), g (1934), a (1876), f (614), c (210), h (69) \\
\hspace{2mm} -- SRA (Semi-regular late-type giants with periods of $35$--$1200$ days) & $45$ & d (40), c (5) \\
\hspace{2mm} -- SRB (More irregular semi-regular late-type giants with periods of $20$--$2300$ days) & $63$ & d (52), c (9), j (2) \\
\hspace{2mm} -- SRC (Semi-regular late-type supergiants) & $2$ & d (2) \\
\hspace{2mm} -- SRD (Semi-regular giants and supergiants of types earlier than SRA, SRB, and SRC) & $1$ & d (1) \\
\hspace{2mm} -- SRS (Semi-regular red giants with shorter periods than other SR types) & $1038$ & d (992), c (46) \\
\hspace{2mm} -- LPV (Unspecified long-period variables) & $9845$ & q (9501), v (166), d (100), c (68), w (7), f (3) \\
PVTEL (PV Telescopii type variables) & $1$ & a (1) \\
RR Lyrae variables & $1057$ & \\
\hspace{2mm} -- ARRD (Anomalous double-mode RR Lyrae variables) & $2$ & y (2) \\
\hspace{2mm} -- RRAB (RR Lyrae variables with asymmetric light curves) & $369$ & a (205), y (94), d (24), b (19), aa (14), q (12), c (1) \\
\hspace{2mm} -- RRC (RR Lyrae variables with symmetric sinusoidal light curves and shorter periods) & $451$ & a (399), d (19), y (19), b (4), c (4), q (3), aa (2), g (1) \\
\hspace{2mm} -- RRD (Double-mode RR Lyrae variables) & $5$ & a (3), q (1), y (1) \\
\hspace{2mm} -- RR (Unspecified RR Lyrae variables) & $230$ & z (221), d (8), c (1) \\
L (Slow irregular variables) & $9771$ & d (9247), a (436), c (84), j (4) \\
\bottomrule
\end{tabular*}
\end{sidewaystable}
\clearpage
\begin{sidewaystable}[t]
\addtocounter{table}{-1}\caption{\enspace Continued.}\label{tab:gds_all_var3}
\centering
\begin{tabular*}{\textheight}{lrl}
\toprule
\textbf{Variability type} & \multicolumn{1}{c}{\textbf{No.}} & \textbf{References} \\
\midrule
SPB (Slowly pulsating B stars) & $11$ & d (10), c (1) \\
PULS (Unspecified pulsating variables) & $729$ & d (729) \\
\midrule
\textbf{Eruptive variability types} & $1459$ & \\
BE (Be variables) & $431$ & a (258), c (129), d (31), ag (8), h (5) \\
cPNB[e] (Compact proto-planetary and planetary nebulae B[e] stars) & $1$ & d (1) \\
DPV (Double periodic variables) & $1$ & c (1) \\
I (Irregular variables) & $8$ & d (5), c (3) \\
RCB (R Coronae Borealis variables) & $25$ & d (12), a (8), c (5) \\
SDOR (S Doradus variables) & $9$ & d (8), c (1) \\
WR (Wolf--Rayet variables) & $13$ & a (6), c (5), d (2) \\
Young stellar objects & $919$ & \\
\hspace{2mm} -- CTTS (Classical T Tauri stars) & $4$ & c (3), ah (1) \\
\hspace{2mm} -- TTS (Unspecified T Tauri stars) & $7$ & ah (4), c (3) \\
\hspace{2mm} -- UXOR (UX Orionis variables) & $4$ & c (2), d (2) \\
\hspace{2mm} -- YSO (Unspecified young stellar objects) & $904$ & a (737), d (94), c (37), ah (36) \\
ERUPTIVE (Unspecified eruptive variables) & $52$ & d (52) \\
\midrule
\textbf{Cataclysmic variability types} & $52$ & \\
Symbiotic variables & $42$ & \\
\hspace{2mm} -- ZAND (Z Andromedae variables) & $14$ & d (11), a (3) \\
\hspace{2mm} -- SYST (Unspecified symbiotic variables) & $28$ & aj (16), ak (12) \\
CV (Unspecified cataclysmic variables, including novae) & $10$ & d (14), a (6), c (2), ai (2) \\
\midrule
\textbf{X-Ray variability types} & $5$ & \\
HMXB (High-mass X-ray binaries) & $4$ & a (2), c (1), d (1) \\
LMXB (Low-mass X-ray binaries) & $1$ & d (1) \\
\midrule
\textbf{Other object types} & $15$ & \\
MICROLENSING (Microlensing events) & $2$ & am (1), an (1) \\
QSO (Quasi-stellar extragalactic sources) & $6$ & al (6) \\
RAD\_VEL\_VAR (Radial velocity variables) & $4$ & ao (4) \\
SB (Spectroscopic binaries) & $3$ & ap (3) \\
\bottomrule
\end{tabular*}
\end{sidewaystable}
\twocolumn

\onecolumn\begin{sidewaystable}[t]
\caption{\enspace Distribution of variable GDS sources over spectral class and main variability group.}\label{tab:gds_types_spt}
\centering
\begin{tabular*}{\textheight}{lccccccccccccc}
\toprule
 & \textbf{O} & \textbf{OB}& \textbf{B} & \textbf{A} & \textbf{F} & \textbf{G} & \textbf{K} & \textbf{M, S, C} & \textbf{WR} & \textbf{Known SpC} & \textbf{Unknown SpC} \\
\midrule
Eclipsing & $100 \,|\, 11$ & $\,\,\,84 \,|\, 0$ & $1924 \,|\, 696\,\,\,$ & $1813 \,|\, 11$ & $1201 \,|\, 0$ & $\,\,\,350 \,|\, 0$ & $\,\,\,424 \,|\, 5$ & $\,\,\,\,\,\,\,\,\,44 \,|\, 0$ & $\,\,\,5 \,|\, 0$ & $5945 \,|\, 723$ & $7086$ \\
Rotating & $\,\,\,23 \,|\, 3\,\,\,$ & $\,\,\,24 \,|\, 0$ & $393 \,|\, 90\,\,\,$ & $108 \,|\, 4$ & $\,\,\,171 \,|\, 0$ & $\,\,\,406 \,|\, 0$ & $1120 \,|\, 9$ & $\,\,\,\,\,\,\,\,\,89 \,|\, 0$ & $\,\,\,0 \,|\, 0$ & $2334 \,|\, 106$ & $1348$ \\
Pulsating & $546 \,|\, 12$ & $\,\,\,52 \,|\, 0$ & $459 \,|\, 115$ & $493 \,|\, 0$ & $\,\,\,418 \,|\, 0$ & $\,\,\,220 \,|\, 0$ & $5844 \,|\, 6$ & $21699 \,|\, 0$ & $11 \,|\, 0$ & $29742 \,|\, 133\,\,\,$ & $19294$ \\
Eruptive & $\,\,\,29 \,|\, 5\,\,\,$ & $\,\,\,85 \,|\, 0$ & $379 \,|\, 73\,\,\,$ & $\,\,\,56 \,|\, 0$ & $\,\,\,\,\,\,18 \,|\, 0$ & $\,\,\,\,\,\,24 \,|\, 0$ & $\,\,\,\,\,\,74 \,|\, 1$ & $\,\,\,\,\,\,\,\,\,40 \,|\, 0$ & $13 \,|\, 0$ & $718 \,|\, 79$ & $662$ \\
Cataclysmic & $\,\,\,\,\,\,0 \,|\, 0\,\,\,$ & $\,\,\,\,\,\,0 \,|\, 0$ & $3 \,|\, 1$ & $\,\,\,\,\,\,0 \,|\, 0$ & $\,\,\,\,\,\,\,\,\,2 \,|\, 0$ & $\,\,\,\,\,\,\,\,\,2 \,|\, 0$ & $\,\,\,\,\,\,\,\,\,2 \,|\, 0$ & $\,\,\,\,\,\,\,\,\,34 \,|\, 0$ & $\,\,\,1 \,|\, 0$ & $44 \,|\, 1$ & $7$ \\
X-ray & $\,\,\,\,\,\,2 \,|\, 0\,\,\,$ & $\,\,\,\,\,\,0 \,|\, 0$ & $2 \,|\, 0$ & $\,\,\,\,\,\,0 \,|\, 0$ & $\,\,\,\,\,\,\,\,\,0 \,|\, 0$ & $\,\,\,\,\,\,\,\,\,0 \,|\, 0$ & $\,\,\,\,\,\,\,\,\,0 \,|\, 0$ & $\,\,\,\,\,\,\,\,\,\,\,\,1 \,|\, 0$ & $\,\,\,0 \,|\, 0$ & $\,\,\,5 \,|\, 0$ & $0$ \\
Other object types & $\,\,\,2 \,|\, 0$ & $\,\,\,\,\,\,1 \,|\, 0$ & $0 \,|\, 0$ & $\,\,\,\,\,\,1 \,|\, 0$ & $\,\,\,\,\,\,\,\,\,1 \,|\, 0$ & $\,\,\,\,\,\,\,\,\,0 \,|\, 0$ & $\,\,\,\,\,\,\,\,\,0 \,|\, 1$ & $\,\,\,\,\,\,\,\,\,\,\,\,0 \,|\, 0$ & $\,\,\,2 \,|\, 0$ & $\,\,\,7 \,|\, 1$ & $7$ \\
Known variability type & $702 \,|\, 31$ & $246 \,|\, 0$ & $3159 \,|\, 975\,\,\,$ & $2471 \,|\, 15$ & $1811 \,|\, 0$ & $1002 \,|\, 0$ & $\,\,\,7462 \,|\, 22$ & $21905 \,|\, 0$ & $32 \,|\, 0$ & $38790 \,|\, 1043$ & $28404$ \\
Unknown variability type & $262 \,|\, 37$ & $299 \,|\, 0$ & $2584 \,|\, 1061$ & $1084 \,|\, 25$ & $\,\,\,412 \,|\, 0$ & $\,\,\,309 \,|\, 0$ & $11662 \,|\, 17$ & $\,\,\,5220 \,|\, 0$ & $16 \,|\, 0$ & $21848 \,|\, 1140$ & $23103$ \\
\midrule
Known variables & $925 \,|\, 49$ & $468 \,|\, 0$ & $4553 \,|\, 1448$ & $2948 \,|\, 22$ & $2036 \,|\, 0$ & $1162 \,|\, 0$ & $15468 \,|\, 36$ & $26225 \,|\, 0$ & $44 \,|\, 0$ & $53829 \,|\, 1555$ & $40327$ \\
New variables & $\,\,\,39 \,|\, 19$ & $\,\,\,77 \,|\, 0$ & $1190 \,|\, 588\,\,\,$ & $\,\,\,607 \,|\, 18$ & $\,\,\,187 \,|\, 0$ & $\,\,\,149 \,|\, 0$ & $3656 \,|\, 3$ & $\,\,\,\,\,\,900 \,|\, 0$ & $\,\,\,4 \,|\, 0$ & $6809 \,|\, 628$ & $11180$ \\
\bottomrule
\end{tabular*}
\begin{tablenotes}
\item Notes: Spectral classifications are drawn from SIMBAD, the VSX, and Gaia~DR3 effective temperatures (left value); additional estimations from GDS $UBV$ photometry (right value) are listed separately. The column for known spectral class (SpC) indicates the sum over all specified spectral classes, while sources with unknown spectral class or a spectral classification flagged as uncertain are summarised as sources with unknown spectral class without differentiation between stellar and non-stellar sources. Other object types include quasi-stellar extragalactic sources, microlensing events, and radial velocity variables. The number of sources with known variability type is herein equated with the sum of all sources with at least one assigned variability type in the catalogue by \citealt{2023A&A...674A..22G} or in the VSX among the variability types listed in Table~\ref{tab:gds_all_var1}. Sources for which variability has been detected in the GDS data or which have been recorded as variable in the VSX, but the variability type is unspecified, either because the sources are unstudied or irregular, are grouped under unknown variability type. Known variables are defined as sources with assignment of at least one variability type in the catalogue by \citealt{2023A&A...674A..22G} or in the VSX including variable stars of unspecified type. Sources that passed the variability criteria detailed in Paper II and are not listed as variables in the catalogue by \citealt{2023A&A...674A..22G} and the VSX are declared as new variables. As mentioned in Section~\ref{Sec:VarTypes}, the variables from the rotating main group should be treated with caution.
\end{tablenotes}
\end{sidewaystable}
\twocolumn

\begin{sidewaystable}[t]
\caption{\enspace $i'$ amplitude ranges in magnitudes of variable stars by spectral class across periodic variability groups of the eclipsing, rotating, and pulsating types; based on the number of stars given in brackets.}\label{tab:gds_amp}
\centering{
\begin{tabular}{lccccccccc}
\toprule
 & \textbf{O} & \textbf{OB}& \textbf{B} & \textbf{A} & \textbf{F} & \textbf{G} & \textbf{K} & \textbf{M, S, C} & \textbf{WR} \\
\midrule
Eclipsing & $0.0$ \textrm{--} $1.4$ & $0.0$ \textrm{--} $1.8$ & $0.0$ \textrm{--} $3.0$ & $0.0$ \textrm{--} $2.6$ & $0.0$ \textrm{--} $2.5$ & $0.0$ \textrm{--} $1.8$ & $0.0$ \textrm{--} $3.0$ & $0.0$ \textrm{--} $1.3$ & $0.1$ \textrm{--} $0.5$ \\
& (100) & (84) & (1920) & (1810) & (1197) & (349) & (421) & (43) & (5) \\
Rotating & $0.0$ \textrm{--} $0.9$ & $0.0$ \textrm{--} $0.7$ & $0.0$ \textrm{--} $1.2$ & $0.0$ \textrm{--} $2.2$ & $0.0$ \textrm{--} $1.8$ & $0.0$ \textrm{--} $1.5$ & $0.0$ \textrm{--} $1.8$ & $0.0$ \textrm{--} $1.3$ & \textrm{--} \\
& (23) & (24) & (393) & (108) & (171) & (405) & (1116) & (89) & (0) \\
Pulsating & $0.0$ \textrm{--} $6.7$ & $0.0$ \textrm{--} $0.9$ & $0.0$ \textrm{--} $2.8$ & $0.0$ \textrm{--} $1.7$ & $0.0$ \textrm{--} $2.8$ & $0.0$ \textrm{--} $1.6$ & $0.0$ \textrm{--} $3.8$ & $0.0$ \textrm{--} $6.7$ & $0.1$ \textrm{--} $1.4$ \\
& (538) & (51) & (457) & (491) & (417) & (219) & (5838) & (21584) & (11) \\
\bottomrule
\end{tabular}
}
\end{sidewaystable}

\begin{sidewaystable}[t]
\caption{\enspace Period ranges in days of variable stars by spectral class across periodic variability groups of the eclipsing, rotating, and pulsating types; based on the number of stars given in brackets.}\label{tab:gds_P}
\centering{
\begin{tabular}{lccccccccc}
\toprule
 & \textbf{O} & \textbf{OB}& \textbf{B} & \textbf{A} & \textbf{F} & \textbf{G} & \textbf{K} & \textbf{M, S, C} & \textbf{WR} \\
\midrule
Eclipsing & $0.2$ \textrm{--} $23.2\,\,\,\,\,\,$ & $0.5$ \textrm{--} $86.6\,\,\,\,\,\,$ & $0.1$ \textrm{--} $176.0\,\,\,$ & $0.0$ \textrm{--} $57.8\,\,\,$ & $0.1$ \textrm{--} $63.2\,\,\,$ & $0.1$ \textrm{--} $103.3\,\,\,$ & $0.1$ \textrm{--} $269.4\,\,\,$ & $0.2$ \textrm{--} $821.3\,\,\,\,\,\,$ & $1.6$ \textrm{--} $39.0\,\,\,$ \\
& (100) & (84) & (1923) & (1812) & (1201) & (350) & (423) & (44) & (5) \\
Rotating & $3.3$ \textrm{--} $18.2\,\,\,\,\,\,$ & $0.5$ \textrm{--} $258.0\,\,\,$ & $0.0$ \textrm{--} $1554.7$ & $0.0$ \textrm{--} $149.8$ & $0.0$ \textrm{--} $89.9\,\,\,$ & $0.1$ \textrm{--} $427.0\,\,\,$ & $0.0$ \textrm{--} $1308.3$ & $0.3$ \textrm{--} $149.8\,\,\,\,\,\,$ & $\textrm{--}$ \\
& (2) & (7) & (208) & (82) & (53) & (90) & (303) & (74) & (0) \\
Pulsating & $0.1$ \textrm{--} $3488.0$ & $0.1$ \textrm{--} $5499.9$ & $0.0$ \textrm{--} $2551.3$ & $0.0$ \textrm{--} $522.2$ & $0.0$ \textrm{--} $487.0$ & $0.0$ \textrm{--} $2571.1$ & $0.1$ \textrm{--} $5495.0$ & $0.1$ \textrm{--} $88874.6$ & $0.4$ \textrm{--} $529.9$ \\
& (446) & (32) & (266) & (211) & (280) & (193) & (4564) & (12275) & (9) \\
\bottomrule
\end{tabular}
}
\end{sidewaystable}

\renewcommand{\thefigure}{A\arabic{figure}}
\begin{figure}[h]
\begin{center}
	\centerline{\includegraphics[width=1.0\columnwidth]{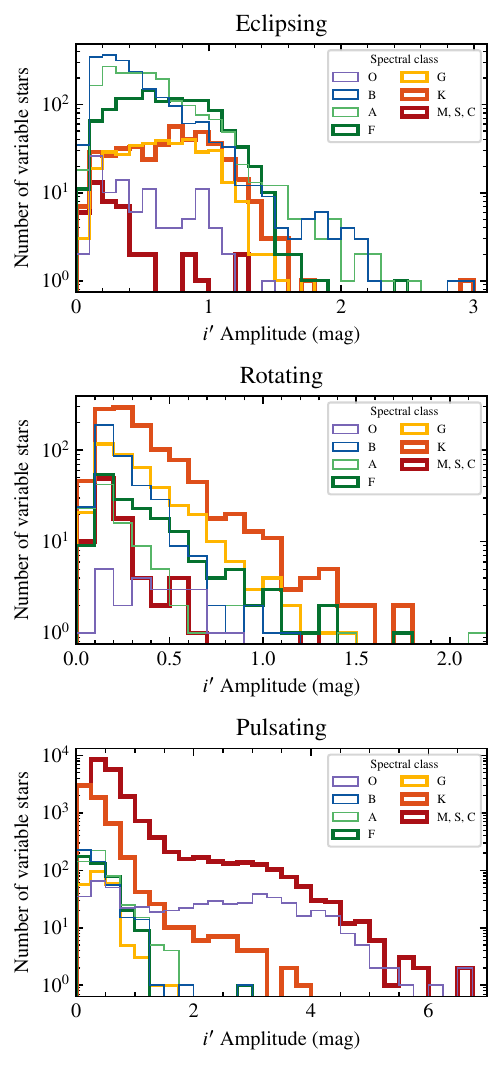}}
	\caption{\enspace $i'$ amplitude distribution of variable stars by spectral class across periodic variability groups of the eclipsing, rotating, and pulsating types. \label{fig:GDS_i_amp_OBAFGKM_ERP}}
\end{center}
\end{figure}
\begin{figure}[h]
\begin{center}
	\centerline{\includegraphics[width=1.0\columnwidth]{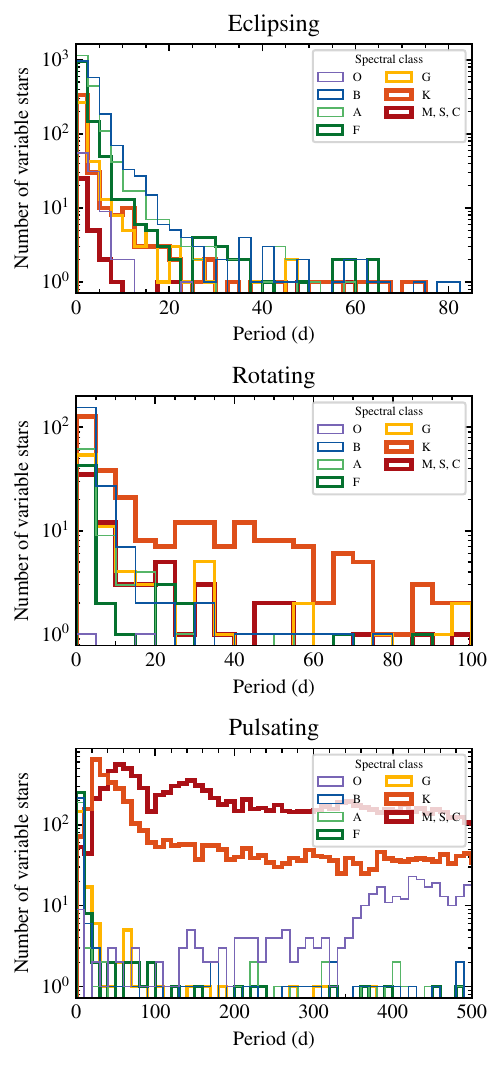}}
	\caption{\enspace Partial view of the period distribution of variable stars by spectral class across periodic main variability groups of the eclipsing, rotating, and pulsating types. \label{fig:GDS_P_OBAFGKM_ERP}}
\end{center}
\end{figure}
\end{document}